\documentclass[12pt]{article}

\usepackage{amsmath,amsthm, amsfonts, amssymb, amsxtra, amsopn}
\usepackage{pgfplots}
\usepgfplotslibrary{colorbrewer}
\pgfplotsset{compat = 1.15, 
			 cycle list/Set1-8} 
\usetikzlibrary{pgfplots.statistics, pgfplots.colorbrewer} 
\usepackage{pgfplotstable}
\usepackage{graphicx,grffile}
\usepackage{multirow}
\usepackage{algorithmic}
\usepackage{booktabs}
\usepackage{listings}
\usepackage{cmap}
\usepackage{colortbl}
\usepackage{adjustbox}
\usepackage{epsfig}
\usepackage{enumerate}
\usepackage{bm}

\usepackage[tableposition=top,font=small,skip=5pt,width=\textwidth]{caption}
\usepackage{subcaption}
\usepackage{makecell} 

\usepackage[explicit]{titlesec}

\PassOptionsToPackage{hyphens}{url}

\usepackage{hyperref}
\hypersetup{colorlinks=true,linkcolor=black,citecolor=black,urlcolor=blue,filecolor=black}
\hypersetup{pdfpagemode=UseNone,pdfstartview=}

\definecolor{darkgreen}{rgb}{0.125,0.5,0.169}

\usepackage{enumitem}
\setlist[itemize]{noitemsep, topsep=0pt}

\advance\oddsidemargin by -0.35in
\advance\textwidth by 0.7in

\advance\topmargin by -0.45in
\advance\textheight by 0.9in

\long\def\symbolfootnotetext[#1]#2{\begingroup%
\def\thefootnote{\fnsymbol{footnote}}\footnotetext[#1]{#2}\endgroup}

\newcommand\dunderline[3][-1pt]{{%
  \sbox0{#3}%
  \ooalign{\copy0\cr\rule[\dimexpr#1-#2\relax]{\wd0}{#2}}}}
\def\uuu{\kern-1pt\dunderline{0.75pt}{\phantom{M}}}

\DeclareMathOperator{\MLP}{MLP}

\DeclareMathOperator*{\argmax}{arg\,max}
\DeclareMathOperator{\logits}{logits}
\DeclareMathOperator{\rgb}{rgb}
\DeclareMathOperator{\grayscale}{gs}
\DeclareMathOperator{\ReLU}{ReLU}
\DeclareMathOperator{\pos}{pos}
\DeclareMathOperator{\nneg}{neg}

\def\Agensla{\texttt{Agensla}}
\def\Androm{\texttt{Androm}}
\def\Convagent{\texttt{Convagent}}
\def\Crypt{\texttt{Crypt}}
\def\Crysan{\texttt{Crysan}}
\def\DCRat{\texttt{DCRat}}
\def\Injuke{\texttt{Injuke}}
\def\Makoob{\texttt{Makoob}}
\def\Mokes{\texttt{Mokes}}
\def\Noon{\texttt{Noon}}
\def\Remcos{\texttt{Remcos}}
\def\Seraph{\texttt{Seraph}}
\def\SnakeLogger{\texttt{SnakeLogger}}
\def\Stealerc{\texttt{Stealerc}}
\def\Strab{\texttt{Strab}}
\def\Taskun{\texttt{Taskun}}
\def\Zenpak{\texttt{Zenpak}}

\def\zz{\phantom{0}}

\title{Neural Fingerprints for Malware Analysis: An Image-Based Metric Learning Approach with Application
to Cross-Domain Classification}

\author{Manasa Deshagouni\footnotemark[1]\ \ \ 
Sayma Akther\footnotemark[1]\ \ \ 
Martin  Jure\v{c}ek\footnotemark[2]\ \ \
Mark Stamp\footnotemark[1]\,\,\footnotemark[3]} 

\begin{document}

\symbolfootnotetext[1]{Department of Computer Science, San Jose State University}
\symbolfootnotetext[2]{Faculty of Information Technology, Czech Technical University in Prague}
\symbolfootnotetext[3]{mark.stamp$@$sjsu.edu}

\maketitle

\abstract
Identifying the family of a newly observed malware sample is a core task in
threat intelligence, yet conventional classifiers must be retrained whenever a
new family appears. This chapter develops an image-based metric learning
approach that instead learns to extract discriminative neural
fingerprints---fixed-length embeddings---from malware-as-image representations,
so that family membership can be determined by nearest-neighbor search in the
embedding space. The central advantage of this formulation is zero-shot
capability: because the learned embedding induces a similarity metric
rather than a fixed set of class boundaries, families that were never seen during
training can be recognized by comparison against a gallery, with no retraining.
We demonstrate this directly by training an encoder on
MalNet-Images-Tiny and MalImg combined (453 families, 96{,}769 images) and
evaluate it zero-shot on a held-out~17-family grayscale dataset with no
family overlap. Using a lightweight CNN with multi-proxy anchor loss, this model
attains~73.1\%\ \texttt{retrieval@1} and~90.5\%\ open-set AUROC---more
than a~12$\times$ improvement over the random baseline---on families the encoder has never seen.
We benchmark our embedding approach against two
conventional paradigms in a same-domain setting, where all three are competitive
at classifying malware into families
(SVM with PCA:~86\%, ResNet-18 classifier:~91\%, metric learning:~94\%, on MalImg).
We further show that the learned embeddings transfer across datasets
(88.5\%\ \texttt{retrieval@1} when trained on MalNet-Images-Tiny and tested on
MalImg without retraining). Unlike classifiers, our embedding approach also
yields interpretable similarity scores and scales to large galleries via
Facebook AI Similarity Search (FAISS). Finally,
we provide a comprehensive evaluation of the learned embedding space using
\texttt{retrieval@k}, cluster purity, silhouette score, separation ratio, few-shot
accuracy, and open-set detection metrics, along with robustness analysis under
image perturbations.

\bigskip



\section{Introduction}

Malware detection is a critical challenge in cybersecurity. As cyber threats evolve,
security analysts face an ever-growing volume of malicious software with large numbers of new
samples appearing daily. Traditional signature-based detection methods, which rely on
byte-sequence matching, are easily evaded through obfuscation, packing, polymorphism, and metamorphism.
These evasion techniques fundamentally alter the binary representation of malware
while preserving malicious functionality, rendering static signatures ineffective.

The problem extends beyond detection to classification, i.e., identifying which malware family a sample belongs to.
Family classification is critical for
incident response (different families require different removal procedures),
threat intelligence (understanding family relationships to track attack distribution and strategies),
risk assessment (some families are more security-critical than others), and
similarity search (finding similar samples to better understand variants and evolution patterns).

Recent advances in deep learning have shown promise for learning robust representations directly
from malware binaries. Although many binary-to-image conversion techniques
have been considered in the literature~\cite{hs25},
typically binary files are converted to grayscale images by mapping
each byte to a pixel value~\cite{Prat}. Such malware images preserve relevant structure,
including PE headers, section boundaries, code regions, and entropy patterns (e.g., encrypted or
packed regions that appear as high-entropy noise).

Visual representations enable the use of advanced Convolutional Neural Network (CNN)
architectures to learn
features automatically, without manual feature engineering. The crucial insight
is that visual similarity in the image space correlates with family relationships, as demonstrated
in numerous research papers~\cite{Niket,Atharva,nataraj2011malware,Huy,Sravani}.

The typical approach to malware classification uses supervised classification,
where a model learns to predict one of~$C$ known families (i.e., classes). However,
classification has some limitations.
For example, classifiers require retraining when new families emerge, which may be impractical
when dealing with rapidly evolving malware.
Classification also generally only provides a class label,
not a measure of how similar a sample is to known families.
And since classifiers learn decision boundaries specific to the training data,
they have limited cross-domain generalizability.
Finally, as the number of families grows, classification becomes increasingly complex,
requiring larger models and more training data.

Metric learning addresses these limitations by learning a distance function in an embedding space,
rather than decision boundaries. An advantage of metric learning is that the embedding space
allows new families to be added without the need to retrain. Another advantage
of metric learning is that similarity scores are obtained, which provides interpretability via a measure of relatedness.
This approach can be viewed as a geometric technique that aligns more closely with how a security analyst
thinks about malware relationships.

The primary goal of the research presented in this chapter is to learn an embedding extractor for malware
images, specifically, a neural fingerprinting model that maps a binary's image representation to a
compact vector such that samples from the same family cluster together.
We then use the resulting neural fingerprints
for zero-shot family recognition, in which families absent from the training set
are identified by nearest-neighbor search against a gallery rather than by a retrained
classifier. To establish how this embedding-based approach compares with conventional
practice, we benchmark it against two standard paradigms. Thus, we compare the
following three approaches.
\begin{enumerate}
\item Classic machine learning (ML) using Support Vector Machines (SVM)
on flattened pixel vectors, with Principal Component Analysis (PCA) for
dimensionality reduction.
\item Deep neural network classification using ResNet-18 with supervised cross-entropy loss.
\item Metric (embedding) learning using ResNet-18 and a lightweight CNN with triplet loss,
progressively refined to multi-proxy anchor loss.
\end{enumerate}
In a closed-set, same-domain setting all three of these approaches are reasonably competitive
(SVM:~86\%, ResNet-18:~91\%, metric learning:~94\%, on MalImg). The decisive difference
emerges out of distribution. Only the embedding model extends to cross-domain transfer
(88.5\%\ \texttt{retrieval@1} when trained on MalNet-Images-Tiny and tested on MalImg)
and, most importantly, to strict zero-shot recognition of unseen families
(73.1\%\ \texttt{retrieval@1} and~90.5\%\ open-set AUROC on a held-out 17-family
grayscale dataset), neither of which a fixed-class classifier can do without retraining.

The contributions of the research presented in this chapter include the following.
\begin{itemize}
\item \textbf{A zero-shot malware family recognition method based on learned image
embeddings} --- Our central contribution is a neural fingerprinting encoder, trained with
multi-proxy anchor loss on MalNet-Images-Tiny and MalImg combined. This approach
recognizes entirely unseen malware families by gallery retrieval---achieving~73.1\%\
\texttt{retrieval@1} and~90.5\%\ open-set AUROC on a disjoint 17-family grayscale test
set, with confirmed zero data leakage and no retraining on the target families.
\item \textbf{A characterization of the learned embedding space} --- We analyze the extracted
family embeddings using \texttt{retrieval@k}, cluster purity, silhouette score, separation
ratio, few-shot accuracy, open-set AUROC, a per-family confusion analysis, and robustness
under image perturbations, giving a multi-faceted view of embedding quality.
\item \textbf{A controlled comparison against conventional paradigms} --- We benchmark our
embedding extractor against an SVM-with-PCA baseline and a supervised ResNet-18 classifier
under a same-domain protocol. We then quantify how performance degrades along the
same-domain $\rightarrow$ cross-domain $\rightarrow$ zero-shot
progression, isolating the out-of-distribution regime where the embedding approach
is uniquely applicable.
\end{itemize}

The remainder of this chapter is organized as follows. In Section~\ref{sect:RW} we briefly discuss
relevant examples of related work. Then in Section~\ref{sect:meth}, we outline our methodology,
with the main focus being our metric learning based malware neural fingerprinting architecture.
Section~\ref{sect:data} contains information on the datasets used in our experiments,
while Section~\ref{sect:exp} provides our experimental results.
In Section~\ref{sect:diss} we discuss the main results and limitations of our work.
Section~\ref{sect:conc} concludes the chapter, and we consider possible directions
for future work.


\section{Related Work}\label{sect:RW}

The literature on malware detection, classification, and analysis is vast,
spanning static, dynamic, and hybrid feature extraction, classical machine learning,
and modern deep learning~\cite{Aycock,Anusha}. In this section, we focus on the
three threads most directly related to our work, namely, image-based malware representation,
distance metric and embedding learning (including its application to security),
and cross-domain and zero-shot generalization. We conclude this section
by positioning our contribution relative to this body of previous work.

\subsection{Image-Based Malware Analysis}

Nataraj et al.~\cite{nataraj2011malware} introduced the now-standard technique of
rendering a binary as a grayscale image by interpreting each byte as an~8-bit
pixel intensity and arranging the bytes in row-major order, with the image
width fixed as a function of file size. Their central empirical observation---that
binaries belonging to the same family produce visually similar textures, even
under minor variation---motivated the entire image-based paradigm, including the
present work. Because this transformation requires no disassembly, unpacking, or
execution, it is attractive as a fast, format-agnostic front end.

A substantial body of subsequent work has applied increasingly expressive models to
these representations. Early approaches paired global texture features (e.g., GIST descriptors)
with classical classifiers~\cite{Sravani}, while later work demonstrated that convolutional neural
networks trained end-to-end on malware images outperform hand-crafted
descriptors~\cite{kalash2018deep}. Transfer learning from ImageNet-pretrained
backbones has been shown to be effective despite the domain gap between natural and
malware images~\cite{Niket}. Ensemble and empirical-comparison studies have further
characterized the effect of architecture and training choices for
image-based malware learning~\cite{Prat,vinayakumar2019robust}. Recent work has examined alternative
binary-to-image transformations---including structured encodings such as QR and
Aztec codes~\cite{Atharva} and systematic comparisons of transformation
techniques~\cite{hs25}---as well as generative approaches that synthesize malware
images for augmentation and adversarial analysis~\cite{Huy}. Byte-sequence models that
operate directly on raw bytes rather than images, such as the use of~1-dimensional CNNs~\cite{jain}
and the convolutional
whole-binary model of Raff et al.~\cite{raff2018malware}, provide a complementary
non-image perspective on learning from unprocessed executables. Collectively, this
line of work establishes that visual or byte-level structure carries strong
family-discriminative signal. Our primary contribution is to convert that signal into a
metric embedding rather than a fixed set of class decisions.

\subsection{Metric Learning and Embedding Learning}

The objective of metric learning is to learn a representation in which a simple
distance reflects semantic similarity. Two broad strategies appear in the literature.
The first learns an explicit parameterized distance on a fixed feature space;
the canonical example is Mahalanobis metric learning, in which a positive
semi-definite matrix~$M$ defines the distance~$d_M(u,v) = \sqrt{(u-v)^{\top} M (u-v)}$
which is optimized so that same-class pairs are
contracted and different-class pairs are expanded. In the malware setting,
distance metric learning of this form has been applied to automated malware
detection~\cite{jurecek_dml1} and to improving the classification of malware
families~\cite{jurecek_dml2}.
Rudd et al.~\cite{Rudd} likewise learn metric embeddings for efficient malware
analysis, demonstrating that compact learned representations support fast similarity
queries at scale.

The second strategy, which we adopt, learns a nonlinear embedding
map~$f_\theta$ and then applies a fixed Euclidean distance on the
L2-normalized output vectors; the metric structure is induced entirely
by the learned map rather than by an explicitly parameterized 
distance.\footnote{In some cases, we report cosine similarity for interpretability, but training
and retrieval rankings are computed under Euclidean distance, which is a
true metric---cosine similarity is not, since it fails the triangle inequality.}
This embedding-based formulation
underlies the most successful open-set recognition systems in computer vision.
FaceNet~\cite{schroff2015facenet} learns face embeddings under a triplet loss such
that distances correspond directly to identity similarity, enabling verification and
clustering over identities never seen during training. In person re-identification,
Hermans et al.~\cite{hermans2017defense} show that a batch-hard triplet variant,
which mines the hardest positive and hardest negative within each mini-batch, is both
simpler and more effective than offline triplet sampling.

Proxy-based losses such as
proxy-anchor loss~\cite{kim2020proxy} replace expensive pairwise or triplet sampling
with a small set of learnable class proxies, accelerating convergence and improving
stability. We build on this idea with a multi-proxy variant for our zero-shot
experiments.

We note that the two strategies (explicit metric versus embedding-based) are
closely related, in the sense that an embedding map composed with
a fixed Euclidean distance induces a (data-dependent) metric on the input space.
However, these two approaches
differ in where the learnable capacity resides. To the best of our knowledge, the
embedding-based formulation with image inputs has not previously been applied to
malware family analysis, which is the gap that we seek to address in this chapter.

\subsection{Cross-Domain and Zero-Shot Generalization}

A model that performs well on its training distribution but degrades on data from a
different source, time period, or collection methodology is of limited operational
value, since malware corpora differ substantially across these axes. Domain
adaptation techniques aim to reduce this gap. For example, domain-adversarial
training~\cite{ganin2016domain} learns features that are simultaneously
discriminative for the task and invariant to the domain, and generative methods have
been used to synthesize malware images that broaden the training
distribution~\cite{Huy}. A central advantage of embedding-based metric learning in
this context is that it naturally supports zero-shot recognition: because the
learned distance is defined on embeddings rather than on a fixed label set, families
absent from the training data can be recognized by nearest-neighbor search against a
gallery, without retraining. In this chapter we use ``cross-domain generalization''
in the strict sense of evaluating a model on a dataset disjoint from its training
data, with no retraining, fine-tuning, or target-domain adaptation, and we
additionally evaluate strict zero-shot generalization to families that are entirely
unseen during training. Significantly, our experiments quantify how performance degrades along the
progression from~$\mbox{same-domain}$ to~$\mbox{cross-domain}$ to~$\mbox{zero-shot}$.


\section{Methodology}\label{sect:meth}

In this section we first formalize our research problem.
We then provide a detailed discussion of our experimental design.
We also include an overview of the various learning architectures,
loss functions (including triplet loss and multi-proxy anchor loss),
the training procedure, and evaluation techniques.

\subsection{Problem Formulation}

Consider a dataset of malware images in the form~$\mathcal{D} = \{(x_i, y_i)\}_{i=1}^N$,
where each~$x_i$ is a grayscale image and~$y_i \in \{1, \ldots, C\}$ is the corresponding
malware family label. Let~$\mathcal{X} =\{x_i\}_{i=1}^N$.
We consider the following three classification approaches.

\begin{description}
\item[Classic ML]--- Learn an SVM classifier~$\mathcal{S}: \mathcal{X} \rightarrow \{1, \ldots, C\}$
that directly predicts the malware family.
\item[Deep Learning]--- Learn a ResNet-18
classifier~$\mathcal{R}: \mathcal{X} \rightarrow [0,1]^C$
that outputs class probabilities via a softmax function.
\item[Metric Learning]--- Learn an encoder~$f_\theta: \mathcal{X} \rightarrow \mathbb{R}^n$
(where~$\theta$ denotes the learnable parameters of the encoder) that maps
all images to~$n$-dimensional embeddings such that for
a specified distance metric~$d(\cdot, \cdot)$ defined on the embedding space
$$
    d\big(f_\theta(x_i), f_\theta(x_j)\big) < d\big(f_\theta(x_i), f_\theta(x_k)\big)
$$
and for all~$i,j,k$ for which~$y_i = y_j$ and~$y_i \neq y_k$.
\end{description}
Next, we provide an overview of each of these three approaches.

\subsubsection{Classic ML: SVM with PCA}

We flatten each~$224 \times 224$ grayscale image to a~50,176-dimensional vector and
then apply normalization via \texttt{StandardScaler}, which standardizes each feature
(i.e., pixel)~$x^{(j)}$ to~$\tilde{x}^{(j)} = (x^{(j)} - \mu_j)/\sigma_j$, where~$\mu_j$
and~$\sigma_j$ are the per-feature mean and standard deviation estimated on the
training set. This zero-mean, unit-variance standardization prevents features with
naturally larger scales from dominating the principal components and is standard
preprocessing prior to PCA. We then project onto the top~256 principal components,
a dimensionality that retains more than~95\%\ of the cumulative explained variance
on MalImg, while keeping the subsequent SVM training tractable. This choice is
supported empirically---reducing to~64 or~128 components yields noticeably lower
classification accuracy, whereas increasing to~512 components increases training
time without a measurable accuracy gain (see Section~\ref{sect:ablation}).
We thus obtain feature vectors~$\phi(x) \in \mathbb{R}^{256}$ that we use to train
and evaluate the SVM. The SVM uses an RBF kernel~$K(x_i, x_j) = \exp(-\gamma\|\phi(x_i) - \phi(x_j)\|_2^2)$.
Since an SVM is inherently a binary classifier, we adopt a one-vs-rest
multiclass strategy to handle our~$C$-class problem: for each family~$c$,
a separate binary SVM is trained to distinguish samples of family~$c$
from samples of all other families, yielding~$C$ binary classifiers in total.
At inference time, all~$C$ classifiers score a given input and the family
whose classifier produces the highest decision value is selected,
yielding the decision function
$$
    \mathcal{S}(x) = \argmax_c \sum_{i \in S_c} \alpha_i y_i K(x_i, x) + b_c
$$
where~$S_c$ is the set of indices of the support vectors for class~$c$,
$K$ is the RBF kernel, $\alpha_i \geq 0$ are the learned dual coefficients,
and~$b_c$ is the bias (offset) term for class~$c$.
We note that PCA is an unsupervised dimensionality-reduction method that
discards the family labels when selecting projection directions. A supervised
alternative such as Linear Discriminant Analysis (LDA), which selects at
most~$C-1$ projections that maximize the ratio of between-class scatter to
within-class scatter, may yield more discriminative features for this
classification task. We leave a systematic PCA-versus-LDA comparison for
future work.

\subsubsection{Deep Learning: ResNet-18}

The penultimate layer of ResNet-18 can be viewed as a feature vector~$F$ of length~512.
This vector typically serves as input to a final fully connected
layer with~$C$ outputs that correspond to the probabilities (via softmax) of
the~$C$ classes.
We employ ResNet-18 as a feature extractor, replacing the final fully-connected layer with
a~$C$-class linear head. The head consists of a learnable weight matrix~$W \in \mathbb{R}^{C \times 512}$
and a bias vector~$b \in \mathbb{R}^C$, both of which are trained jointly with the backbone.
Given a feature vector~$F(x)$ obtained from sample~$x$ by passing it through the backbone,
the head produces unnormalized class scores~$\logits = W \cdot F(x) + b$, and these
scores are converted to class probabilities through the softmax function
$$
    p(y=c\,|\,x) = \exp(\logits_c) \bigg/ \displaystyle\sum_{j=1}^C \exp(\logits_j) .
$$
The actual output of the model is the probability vector~$\mathcal{R}(x) = (p_1, \ldots, p_C)$,
and~$\logits$ is the intermediate quantity from which probabilities are computed.
We use the term~$\logits$ in the standard \texttt{PyTorch} sense, namely as the unnormalized
linear scores fed into the softmax---these are related to, but not literally equal to,
the log-odds~$\ln\big(p_c / \sum_{j \neq c} p_j\big)$ derived from the final probabilities.
For training, we use the cross-entropy loss~$\mathcal{L} = -\sum_i y_i \log\hat{y}_i$,
where~$y_i$ is the true label for class~$i$ (one-hot encoded)
and~$\hat{y}_i$ is the predicted probability for class~$i$.

\subsubsection{Metric Learning with Triplet and Multi-Proxy Anchor Loss}

Unlike classification techniques, which learn decision boundaries, metric learning yields a distance
metric~$d: \mathbb{R}^n \times \mathbb{R}^n \rightarrow \mathbb{R}^+$ on the embedding space,
where
$$
    d\big(f_\theta(x_i), f_\theta(x_j)\big) < d\big(f_\theta(x_i), f_\theta(x_k)\big)
    	\text{ whenever } y_i = y_j \text{ and } y_i \neq y_k .
$$
This constraint ensures that samples from the same family are closer in the
embedding space implied by~$d(\cdot,\cdot)$ than they are to samples from different families.
A key advantage of metric learning is that the learned metric is universal,
in the sense that it applies to any malware sample, including samples from
families not seen during training. This enables zero-shot capability, where
new families can be identified by finding nearest neighbors in the embedding space
without retraining.

In the experiments reported in Section~\ref{sect:exp}, we use two loss functions
within this metric-learning paradigm: a batch-hard triplet loss for the same-domain
and standard cross-domain settings, and a multi-proxy anchor loss for the
strict zero-shot setting, where the test families are entirely unseen during training.
Both losses are described in Section~\ref{sect:N_finger}.

We now provide additional information on our experimental design. First, we
present our neural fingerprinting architecture, then we discuss the various architectural
design elements in some detail.

\subsection{Neural Fingerprint Architecture}\label{sect:N_finger}

Our encoder architecture consists of three components, namely, a backbone network,
a projection head, and embedding normalization. We discuss each of these three
components in this section.

We employ ResNet-18~\cite{he2016deep} as our default feature extractor.\footnote{A
lightweight ``small'' CNN is also evaluated against ResNet-18 in the ablation study
in Section~\ref{sect:ablation}, where we show that the smaller backbone in fact
generalizes better in our strict zero-shot setting.}
For this task, we chose a Residual Network because
ResNet's skip connections mitigate the vanishing gradient problem,
thereby enabling the training of very deep networks. This is crucial for learning hierarchical
features from malware images, where low-level patterns (e.g., edges)
must be combined into high-level family-specific 
structures.

We initialize our ResNet model with pretrained weights; specifically,
we use the pretrained weights from ImageNet~\cite{deng2009imagenet}.
This provides a strong starting point for fine-tuning the model.
Even though ImageNet contains natural images rather than malware, the early layers learn generic
edge and texture detectors, and previous work indicates that transfer learning
works well for grayscale malware images~\cite{Niket,Sravani}.

ResNet's modular design allows easy adaptation to different input channels.
We adapt the first convolutional layer for single-channel grayscale ($\grayscale$)
input by averaging the pre-trained RGB weights, that is,
$$
    W_{\!\grayscale} = \frac{1}{3}\sum_{c=1}^{3} W_{\!\rgb}^{(c)}
$$
where~$W_{\rgb}^{(c)}$ are the original ImageNet-pretrained weights for channel~$c$.
This averaging preserves the learned filters while adapting to grayscale input,
thus maintaining the benefits of transfer learning.

ResNet-18 provides an excellent efficiency-to-accuracy trade-off.
Our ablation studies, which appear in Section~\ref{sect:ablation}, confirm this in two ways.
First, on the same-domain MalImg dataset, ResNet-18 matches the accuracy of
the deeper ResNet-50~\cite{he2016deep}---a 50-layer variant of the same residual
architecture family---while training in roughly half the time.
In our strict zero-shot setting,
a smaller lightweight CNN actually generalizes better than ResNet-18,
suggesting that model capacity is not the primary bottleneck for malware images,
which are relatively simple compared to natural images.

Our backbone model outputs feature vectors~$F$ of dimension~$n$,
where~$n=512$ for ResNet-18 and~$n=2048$ for ResNet-50.
These are passed through a two-layer MLP projection head to obtain
$$
    z = \MLP(F) = W_2 \cdot \ReLU(W_1 \cdot F + b_1) + b_2
$$
where~$W_1 \in \mathbb{R}^{256 \times n}$, $W_2 \in \mathbb{R}^{d \times 256}$,
$b_1 \in \mathbb{R}^{256}$, $b_2 \in \mathbb{R}^{d}$,
and~$d$ is the desired embedding dimension. We use~$d=128$ for the same-domain
and standard cross-domain experiments, and~$d=256$ for the strict zero-shot setting,
which we found in ablation (Section~\ref{sect:ablation}) to provide slightly better
generalization to unseen families.

The final embeddings are L2-normalized as 
$$
    \hat{z} = \frac{z}{\|z\|_2}
$$
These normalized embeddings lie on the unit hypersphere, which serves the following purposes.
\begin{description}
\item[\textbf{Geometric constraint}]--- Normalizing all embeddings to the surface of a
unit hypersphere prevents embeddings from
growing unbounded during training, which can destabilize metric learning optimization.
\item[\textbf{Efficient similarity computation}]--- For normalized vectors, cosine similarity reduces to a simple dot
product: $\cos(\hat{z}_i, \hat{z}_j) = \hat{z}_i^T \hat{z}_j$. This enables extremely fast similarity search
using FAISS's inner product index, which we discuss in Section~\ref{sect:FAISS}.
\item[\textbf{Scale invariance}]--- Whereas the previous point concerns the geometry
of the embedding space, this point concerns its semantics: by collapsing
embedding magnitudes to~1, the model is forced to encode all family
information in the \emph{direction} of the embedding vector rather than its
length. Magnitudes are sensitive to dataset-specific feature-activation
scales, such as differences in mean image intensity between datasets, 
so removing them improves cross-domain generalization.
\item[\textbf{Training stability}]--- Without normalization, triplet loss can cause embeddings to grow indefinitely
as the model tries to push negatives further apart. Normalization prevents this by constraining the
embedding space, leading to more stable training dynamics.
\item[\textbf{Hypersphere geometry}]--- In high-dimensional spaces, most of the volume of a hypersphere is
concentrated near the surface~\cite{blum2020foundations}.
Thus, by constraining embeddings to the surface, we do not lose much
representational capacity, while gaining the benefits discussed above.
\end{description}

Triplet loss is designed to learn a metric space, where samples from the same class are closer
to each other than samples from different classes. For each sample~$x$, which is referred to as
an anchor, we want
$$
    d(f_\theta(x), f_\theta(x_{\pos})) + \alpha < d(f_\theta(x), f_\theta(x_{\nneg}))
$$
where~$x_{\pos}$ is a positive sample from the same class as~$x$, $x_{\nneg}$
is a negative sample from a different class than~$x$, and~$\alpha$ is a margin that enforces a minimum separation.
This constraint ensures that intra-class distances are smaller than inter-class distances by at least~$\alpha$.

We employ batch-hard triplet loss~\cite{hermans2017defense},
which, for each anchor, selects the hardest positive and the hardest
negative within the current mini-batch. The hardest positive is the
same-class sample that is farthest from the anchor in embedding space,
while the hardest negative is the different-class sample that is closest.
In this context, ``hardest'' means most likely to violate
the margin constraint and therefore most informative for learning.
For a batch of embeddings~$\{z_i\}_{i=1}^B$ with
corresponding labels~$\{y_i\}_{i=1}^B$, the loss is
$$
    \mathcal{L} = \frac{1}{B}\sum_{i=1}^B \max(0, \alpha + d_{i,\pos}^* - d_{i,\nneg}^*)
$$
where
$$
    d_{i,\pos}^* = \max_{\{j\,|\,y_j = y_i, j \neq i\}} \|z_i - z_j\|_2 \quad \text{and}\quad 
    d_{i,\nneg}^* = \min_{\{j\,|\,y_j \neq y_i\}} \|z_i - z_j\|_2 
$$
and~$\alpha$ is the margin hyperparameter.
The ``hardest positive'' (i.e., $d_{i,\pos}^*$) is the positive sample farthest from the anchor,
which can be viewed as the most difficult to ``pull'' into the correct class. The ``hardest negative'' 
(i.e., $d_{i,\nneg}^*$) is the
negative sample closest to the anchor, which can be viewed as the most difficult to ``push'' apart.
Thus, batch-hard triplet loss ensures that the model focuses on the most informative training examples.

Traditionally, triplet loss uses randomly sampled triplets, which can be inefficient
since many triplets likely already satisfy the margin constraint and contribute little to learning.
Batch-hard mining focuses on the hardest triplets, that is, those triplets that most violate the margin
constraint. This provides several benefits, including
faster convergence, better separation, and computational efficiency.
Faster convergence follows from hard triplets providing stronger learning signals,
thereby accelerating convergence. By focusing on difficult cases, the model learns more
discriminative embeddings, thereby yielding better separation.
 With respect to computational efficiency, since all triplets are computed from the same batch,
 efficient GPU utilization is possible.

For our strict zero-shot experiments, where the test families are entirely disjoint from
the training families, we found that batch-hard triplet loss alone was prone to overfitting
to the training family boundaries. We therefore additionally employ multi-proxy anchor
loss~\cite{kim2020proxy}, which represents each training class by~$P$ learnable proxy
vectors and pulls anchors of class~$c$ toward all proxies of class~$c$ while pushing
them away from proxies of other classes. Using multiple proxies per class better captures
intra-class variability, which we hypothesize is what helps the model learn lower-level
visual features---such as code structure and entropy patterns---that, while shaped by the
training families, are general enough to transfer to some unseen families.
We use~$P=4$ proxies per class for the zero-shot experiments, which
is in the range suggested by Kim et al.~\cite{kim2020proxy}. 

For normalized embeddings, we use the cosine-to-Euclidean conversion
$$
    \|z_i - z_j\|_2^2 = 2\big(1 - \cos(z_i, z_j)\big)
$$
where $\cos(z_i, z_j) = z_i^T z_j$ for our normalized vectors. This equivalence allows us to compute
Euclidean distances efficiently using dot products, which are highly optimized in modern hardware.
Note that the factor of~2 comes from the geometry of the unit hypersphere---the maximum Euclidean
distance between two points on a unit sphere is~2, which occurs when they are antipodal.

\begin{figure}[!htb]
\centering
\includegraphics[scale=0.475]{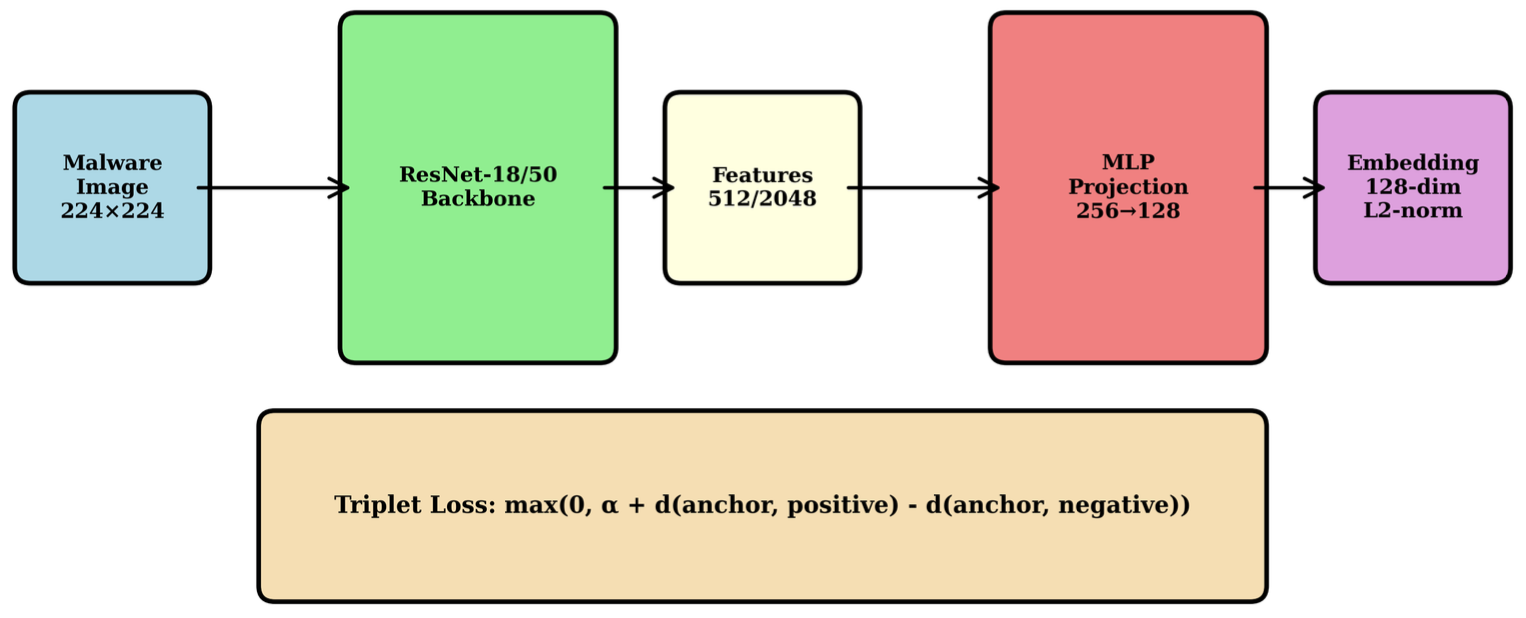}
\caption{Neural fingerprint architecture}\label{fig:architecture}
\end{figure}

To summarize, our neural fingerprint architecture consists of malware images
processed through a CNN backbone (ResNet-18 by default; a lightweight small
CNN in the zero-shot setting), projected onto
embeddings via an MLP, followed by L2-normalization, and optimized via a metric learning loss
(batch-hard triplet loss for same-domain and standard cross-domain experiments;
multi-proxy anchor loss for the strict zero-shot setting).
This architecture is summarized in Figure~\ref{fig:architecture}.

\subsection{Training Procedure}

Training is performed using the components and techniques listed in Table~\ref{tab:train}.
We note that cosine annealing gradually reduces the learning rate based on a cosine curve
$$
    \text{LR}(t) = \text{LR}_0 \cdot \frac{1 + \cos(\pi t / T_{\max})}{2}
$$
where~$t$ is the current epoch and~$T_{\max} = \max(10, \text{epochs})$ is the annealing period.
This schedule provides a high learning rate at the start, thus enabling rapid learning,
with a decreasing learning rate toward the end that enables fine-grained optimization.
Smooth decay prevents oscillations and ensures stable convergence.

\begin{table}[!htb]
    \centering
    \caption{Training components}\label{tab:train}
    \begin{adjustbox}{scale=0.85}
        \begin{tabular}{c|cc}
            \toprule
            \textbf{Description} & \textbf{Hyperparameter} & \textbf{Value} \\
            \midrule
            \multirow{3}{*}{AdamW optimizer} & learning rate & $10^{-4}$ \\
                             & weight decay & $10^{-4}$ \\
                             & regularization & L2 \\ \midrule
              Learning rate scheduler & cosine annealing & $T_{\max} = \max(10, \text{epochs})$ \\ \midrule
               \multirow{2}{*}{Precision} & \multirow{2}{*}{AMP} & FP16  \\
              				& & FP32 (critical) \\ \midrule
              \multirow{4}{*}{Data augmentation} & random resized crop & --- \\
                             & rotation & $\pm 10^{\circ}$ \\
                             & horizontal flip & --- \\
                             & normalization & --- \\ \midrule
              Batch & batch size & 64 \\ \midrule
              \multirow{2}{*}{Embeddings} & dim (same-/cross-domain) & 128 \\
              	  & dim (zero-shot) & 256 \\ \midrule
              ResNet & epochs & 10 to 80 \\ \midrule
              Triplet & margin & 0.2 \\
             \bottomrule
        \end{tabular}
    \end{adjustbox}
\end{table}

On CUDA devices, we employ Automatic Mixed Precision (AMP), which uses FP16
precision for most operations,
with FP32 used for critical operations, such as loss computation. This provides~2$\times$
faster training on modern GPUs, a 2$\times$ reduction in memory usage (enabling larger batch sizes),
while having a minimal impact on final accuracy---typically less than a~0.1\%\ difference~\cite{micikevicius2018mixed}.

Note that data augmentation is used to prevent overfitting and to improve generalization.
This is especially important for the cross-domain scenarios that we consider.

Our ablation studies below show that a batch size of~64 provides
an optimal balance between batch-hard mining effectiveness (i.e., having enough
candidates per anchor for informative hardest-positive and hardest-negative
selection) and computational efficiency.
Furthermore, the embedding dimension of~128 (same-domain and standard cross-domain)
and~256 (strict zero-shot) provides a balance between representational capacity,
computational efficiency, and generalization.

We define our primary evaluation metric \texttt{retrieval@1} in
Section~\ref{sect:FAISS}; informally, it is the proportion of query samples whose
single nearest neighbor in the gallery belongs to the same family.
Validation is performed after each epoch using \texttt{retrieval@1}
accuracy on the validation set.
We checkpoint the model with the highest validation accuracy, ensuring we save the
best-performing model rather than the final model.
This early-stopping strategy reduces overfitting and ensures better performance
on unseen data.

\subsection{Retrieval and Evaluation}\label{sect:FAISS}

\begin{figure}[!htb]
\centering
\includegraphics[width=0.60\textwidth]{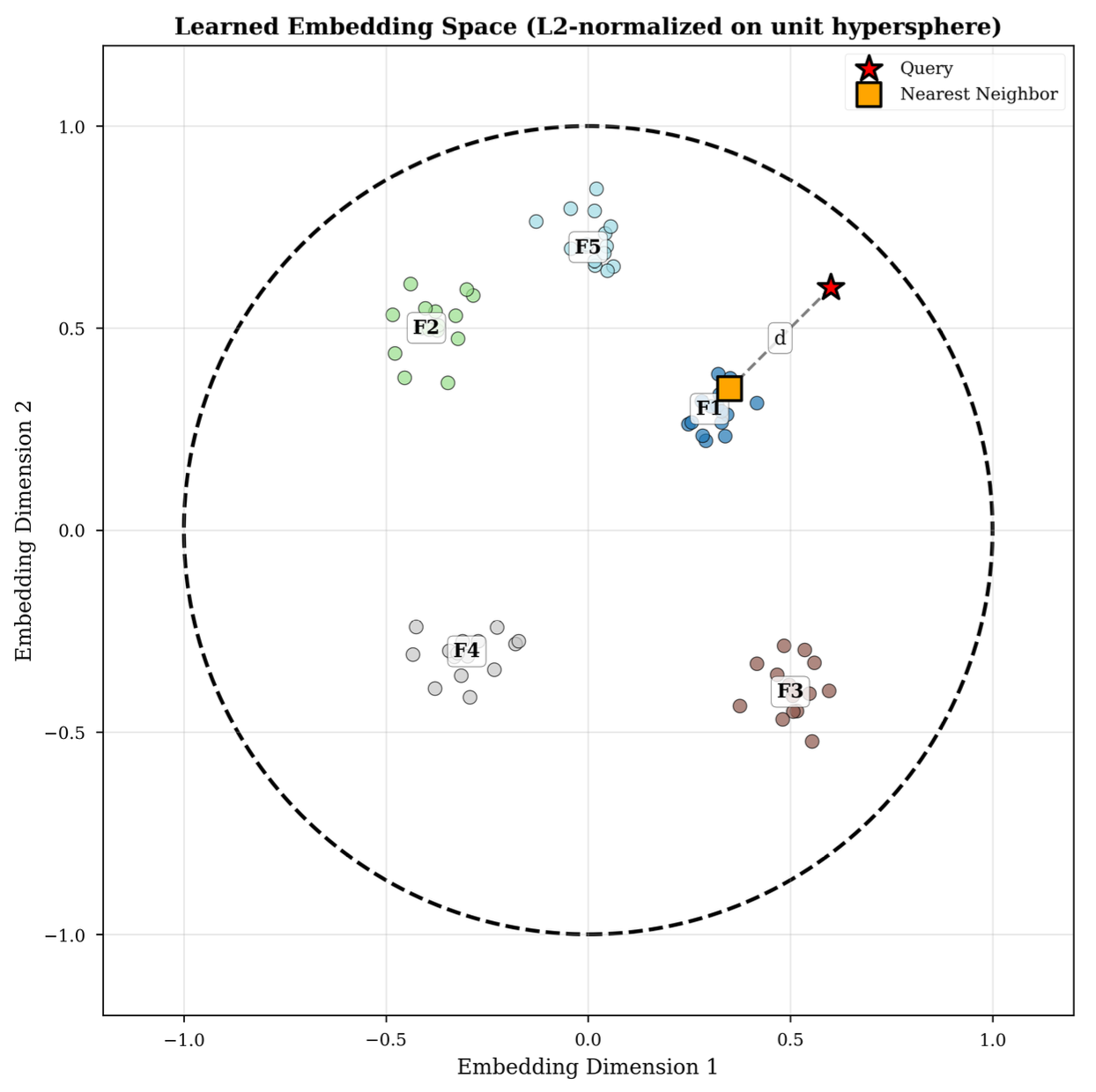}
\caption{Conceptual visualization of the learned embedding space}\label{fig:embedding_space}
\end{figure}

Embeddings are L2-normalized to lie on the unit hypersphere,
with samples from the same malware family ideally forming tight clusters.
Query samples are matched via a nearest-neighbor search.
This clustering and nearest-neighbor
process is illustrated in Figure~\ref{fig:embedding_space}.

As the number of samples grows, brute-force nearest-neighbor search becomes computationally prohibitive.
Facebook AI Similarity Search (FAISS) provides optimized algorithms for similarity search.
For our normalized embeddings, we use \texttt{IndexFlatIP} (inner product index), which computes
cosine similarity via dot products. This is optimal for our use case since all embeddings are L2-normalized,
making the inner product equivalent to cosine similarity.

In real-world deployment, we maintain a ``gallery'' of known malware samples,
stored as indexed embeddings, and process incoming ``queries''---that is,
new, previously unseen samples---by encoding them and searching the gallery.
This gallery/query paradigm enables new families to be added without
retraining, yielding a zero-shot capability.
Also, the gallery can be updated incrementally as new samples arrive.
Since the gallery is indexed once and queried many times, the system is highly scalable.

We consider the following metrics to assess embedding quality,
each of which captures a different aspect of the embedding space.
\begin{description}
\item[\textbf{\texttt{retrieval@$\bm{k}$}}]--- This is our primary metric, which measures the fraction of queries
where the true class appears among the top-$k$ retrieved neighbors.
For this metric, \texttt{retrieval@1} (top-1 accuracy)
is the most stringent, as the nearest neighbor must belong to the correct family. This metric directly
measures the practical utility of the embeddings for malware classification.
Note that for~$k \geq 2$, \texttt{retrieval@$k$} is not equivalent to
$k$-nearest-neighbor classification: \texttt{retrieval@$k$} counts a query as correct if
at least one of the top-$k$ neighbors has the true label, whereas a majority-vote
KNN classifier could still misclassify that same query (e.g., if two of the top three
neighbors belong to other families). We report \texttt{retrieval@$k$} because it directly
characterizes whether a correct match is surfaced to an analyst within the top-$k$ results,
which is a quantity of interest in a retrieval/triage workflow.
\item[\textbf{Cluster purity}]--- We measure the homogeneity of clusters formed by $K$-means clustering.
Specifically, we compute
$$
    \text{Purity} = \frac{1}{N}\sum_{j=1}^{C} \max_i |C_j \cap L_i|
$$
where $C_j$ are clusters, $L_i$ are true class labels, and~$N$ is the total number of samples.
Purity ranges from~0 to~1,
where 1 indicates perfect clustering, that is, each cluster contains samples corresponding to only one family.
\item[\textbf{Silhouette score}]--- This well-known measure is computed for sample~$i$ as
$$
    s(i) = \frac{b(i) - a(i)}{\max\big(a(i), b(i)\big)}
$$
where~$a(i)$ is the mean intra-cluster distance (distance to other samples in the same cluster)
and $b(i)$ is the mean nearest-cluster distance (distance to the nearest cluster).
Silhouette score ranges from~$\hbox{}-1$ to~$\hbox{}+1$, with~$s(i) \approx +1$ being optimal
since the sample is relatively close to samples in its own cluster and relatively far from samples in other
clusters.
\item[\textbf{Separation ratio}]--- This is the ratio of inter-class to intra-class distances
$$
    \text{SepRatio} = \frac{\text{mean inter-class distance}}{\text{mean intra-class distance}}
$$
Higher ratios indicate better separation.
A~\texttt{SepRatio} of~$x$ indicates that, on average, samples
from different families are~$x$ times farther apart than samples from the same family.
\item[\textbf{Open-set AUROC}]--- For the strict zero-shot setting, we additionally report the Area
Under the ROC curve (AUROC) for distinguishing in-gallery queries from out-of-gallery queries,
where the score is the cosine similarity to the nearest gallery sample.
This measures the model's ability to flag samples that do not belong to any known family.
\item[\textbf{$\bm{N}$-way $\bm{K}$-shot accuracy}]--- Also for the zero-shot setting, we sample~$N$ unseen
families and~$K$ labeled examples per family to form a small support set, then classify
query samples by nearest support-set centroid. This evaluates how quickly the embedding
space adapts to new families given very few labeled examples.
\end{description}
Each of these metrics captures a different aspect of the data.
Retrieval accuracy indicates whether a sample from the correct family
is among the nearest neighbors,
cluster purity measures whether families tend to form clusters,
the silhouette score tells us whether individual samples are well-positioned,
and the separation ratio indicates the margin for error. Open-set AUROC and
few-shot accuracy add complementary views relevant when new families appear.
Together, these metrics provide a comprehensive assessment of embedding quality.


\section{Datasets}\label{sect:data}

In this research, we consider three malware image datasets: MalImg, MalNet-Images-Tiny,
and a held-out 17-family grayscale dataset used exclusively for our strict zero-shot evaluation.
Note that none of these datasets was filtered to exclude packed or encrypted samples;
as discussed in Section~\ref{sect:lim}, image-based approaches are less effective for such samples
since packing and encryption destroy the byte-level structure that yields recognizable visual patterns.
The total number of samples in these datasets---and the number of samples used for
training, validation, and testing---are given in Table~\ref{tab:data}.

\begin{table}[!htb]
    \centering
    \caption{Datasets}\label{tab:data}
    \begin{adjustbox}{scale=0.85}
        \begin{tabular}{l|cccc}
            \toprule
            \multicolumn{1}{c|}{\multirow{2}{*}{\raisebox{-1.5pt}{\textbf{Dataset}}}}
            	& \multicolumn{4}{c}{\textbf{Number of samples}} \\ \cmidrule(lr){2-5} 
            					& \textbf{Train} & \textbf{Validation} & \textbf{Test} & \textbf{Total} \\
            \midrule
            MalImg  			& \zz5,669
            					& 1,133
						& \zz2,537
						& \zz9,339 \\
            MalNet-Images-Tiny  	& 61,201
            					& 8,743
						& \zz5,945
						& 75,889 \\
            17-family grayscale (test only) & ---
            					& ---
						& 17,000
						& 17,000 \\ \bottomrule
	\end{tabular}
    \end{adjustbox}
\end{table}

The MalImg dataset~\cite{nataraj2011malware} contains~9,339 malware samples from~25 families,
converted to grayscale images.
The malware families represented include \texttt{Adialer.C}, \texttt{Agent.FYI}, \texttt{Allaple.A}, \texttt{Allaple.L},
and others, with significant class imbalance, ranging from~8 to~2,359 samples per family.
MalNet-Images-Tiny~\cite{freitas2021malnet} is a subset of the larger MalNet dataset. This malware image
dataset contains~75,889 malware images across~47 families. This dataset includes
families representing several types of malware, including adware, Trojans, backdoors,
and spyware. For the remainder of this chapter, we use ``MalNet''
as shorthand for the MalNet-Images-Tiny dataset.

It is worth noting that no universal malware naming standard has achieved
industry-wide adoption, although most antivirus vendors follow variations
of the scheme proposed by the Computer Antivirus Research
Organization~(CARO)~\cite{bontchev_caro}. The CARO Malware Naming Scheme
groups malware into families based on code similarity and uses a
hierarchical format---the version formally adopted by Microsoft, for example,
follows the pattern
\texttt{Type:Platform/Family.Variant!Suffixes}~\cite{caro_netsurion}.
The family labels in the three datasets used in this chapter reflect different
labeling pipelines that are all rooted in this tradition.
MalImg~\cite{nataraj2011malware} was labeled using Microsoft Security
Essentials, whose naming convention is a direct implementation of the CARO
scheme.
MalNet~\cite{freitas2021malnet} obtains its family and type labels from
Euphony~\cite{euphony}, a labeling system that aggregates and reconciles
the output of up to~70 antivirus vendors via VirusTotal; because each
vendor applies its own CARO variant, Euphony acts as a consensus mechanism
over multiple naming conventions.
The~17-family grayscale dataset derives its family labels from the
RawMal-TF dataset~\cite{rawmaltf}, which used ClarAVy~\cite{claravy}
to integrate type-level labels and parsed family information from the
binary names originally assigned by the source repositories.
Because the three datasets were labeled by different tools at different
times, some inconsistency in naming granularity is inevitable; however,
all labels ultimately trace back to antivirus-vendor family assignments
that follow CARO-derived conventions.

The third dataset, which we refer to as the~17-family grayscale dataset, consists
of~17,000 grayscale images drawn from~17 malware families that do not appear
in either MalImg or MalNet. This malware image dataset was generated in~\cite{hs25},
with the corresponding binaries
being from the RawMal-TF dataset~\cite{rawmaltf}.
We use this dataset only as a held-out test set to evaluate strict
zero-shot generalization, that is, generalization to families not seen during training.
The~17 families are
\Agensla, \Androm, \Convagent, \Crypt, \Crysan, \DCRat, \Injuke, \Makoob, \Mokes,
\Noon, \Remcos, \Seraph, \SnakeLogger, \Stealerc, \Strab, \Taskun, and~\Zenpak.
We verified that there is zero overlap in image hashes between this test set and the
combined MalNet+MalImg training pool, so the zero-shot evaluation is free of data leakage.

The MalImg and MalNet images were generated using the methodology of
Nataraj et al.~\cite{nataraj2011malware}, where each byte in the binary is mapped
to a pixel value in~$0$--$255$, bytes are arranged in row-major order,
and the image width is chosen as a function of file size.
This yields variable-size images that summarize the entire binary and require 
no disassembly, unpacking, or execution.
The~17-family grayscale dataset was constructed differently. 
These images were generated from the first~$224\times 224 = 50{,}176$ bytes
of each sample, padding with~0 bytes, if necessary~\cite{hs25}.
Also, the~17-family grayscale images were derived from malware samples in 
the RawMal-TF dataset~\cite{rawmaltf}.
As a result, the 17-family evaluation imposes at least
the following three forms of distribution shift simultaneously on the encoder.
\begin{enumerate}[label=(\roman*)]
\item The families are disjoint from anything seen during training,
\item The source binaries come from a different corpus (RawMal-TF)
than MalNet or MalImg, and
\item The image-construction procedure differs from the
Nataraj-style whole-binary rendering used for the training images. 
\end{enumerate}
We view this combination of shifts as a stringent test of the embedding
extractor and a challenging setting in which to evaluate strict zero-shot
generalization.

All images are resized to~$224 \times 224$ pixels (the standard ImageNet size)
and converted to grayscale.
As mentioned above, during training, we apply data augmentation to improve generalization.
To ensure that retrieval scores are reproducible across runs, augmentation
is not applied during the evaluation phase. Instead, every input image is
zero-padded to a square (preserving its aspect ratio), bilinearly resized
so that its shorter side is~256 pixels, and then center-cropped
to~$224 \times 224$. This deterministic pipeline removes the
stochasticity of random cropping
while keeping the input dimensions consistent
with those expected by the ImageNet-pretrained backbone.


\section{Experiments and Results}\label{sect:exp}

We evaluate our approach across three settings of increasing difficulty. The first two
establish that the learned embeddings are sound and that they transfer across datasets;
the third---strict zero-shot recognition of unseen families---is the central result of
the chapter and the setting that motivates the embedding-based formulation.
\begin{enumerate}
\item \textbf{Same-domain} --- We train and test on the same dataset. In this scenario,
we report results
on MalImg for all three paradigms (SVM, ResNet-18 classifier, and triplet-loss metric
learning) and we also report a same-domain MalNet metric-learning baseline.
\item \textbf{Standard cross-domain} --- We train a triplet-loss encoder on MalNet,
evaluate without any retraining on the MalImg test set.
\item \textbf{Strict zero-shot (primary result)} --- We train on MalNet and MalImg combined,
then evaluate without retraining on the held-out 17-family grayscale dataset, whose families
do not appear in the training data. This setting directly measures the embedding
extractor's ability to fingerprint families it has never seen.
\end{enumerate}
All experiments use the hyperparameters specified in Table~\ref{tab:train} unless
explicitly noted otherwise. The development environment for our experiments
is summarized in Table~\ref{tab:hardSoft}.

\begin{table}[!htb]
    \centering
    \caption{Hardware and software}\label{tab:hardSoft}
    \begin{adjustbox}{scale=0.85}
        \begin{tabular}{c|cc}
            \toprule
            & \textbf{Component} & \textbf{Details} \\
            \midrule
            \multirow{3}{*}{Hardware}
            & CPU & Apple Silicon (M-series) \\
            & Memory & 16GB RAM \\
            & Storage & 10GB (dataset/checkpoints)\\ \midrule
            \multirow{9}{*}{Software}
            & Python & 3.8 (tested on 3.12) \\
            & \texttt{PyTorch} 2.0+ & CPU or CUDA \\
            & \texttt{torchvision} & Image datasets and transforms \\
            & \texttt{numpy} & Array and embedding operations \\
            & \texttt{PIL 7} & Image loading and format conversion \\
            & \texttt{scikit-learn}  & Evaluation metrics \\
            & \texttt{faiss-cpu} or faiss & Similarity search \\
            & \texttt{streamlit} & UI (optional) \\
            & \texttt{matplotlib} & Visualizations \\
            \bottomrule
        \end{tabular}
    \end{adjustbox}
\end{table}

\subsection{Same-Domain Experiments on MalImg}\label{sect:exp:samedomain}

\begin{figure}[!htb]
\centering
\includegraphics[scale=0.45]{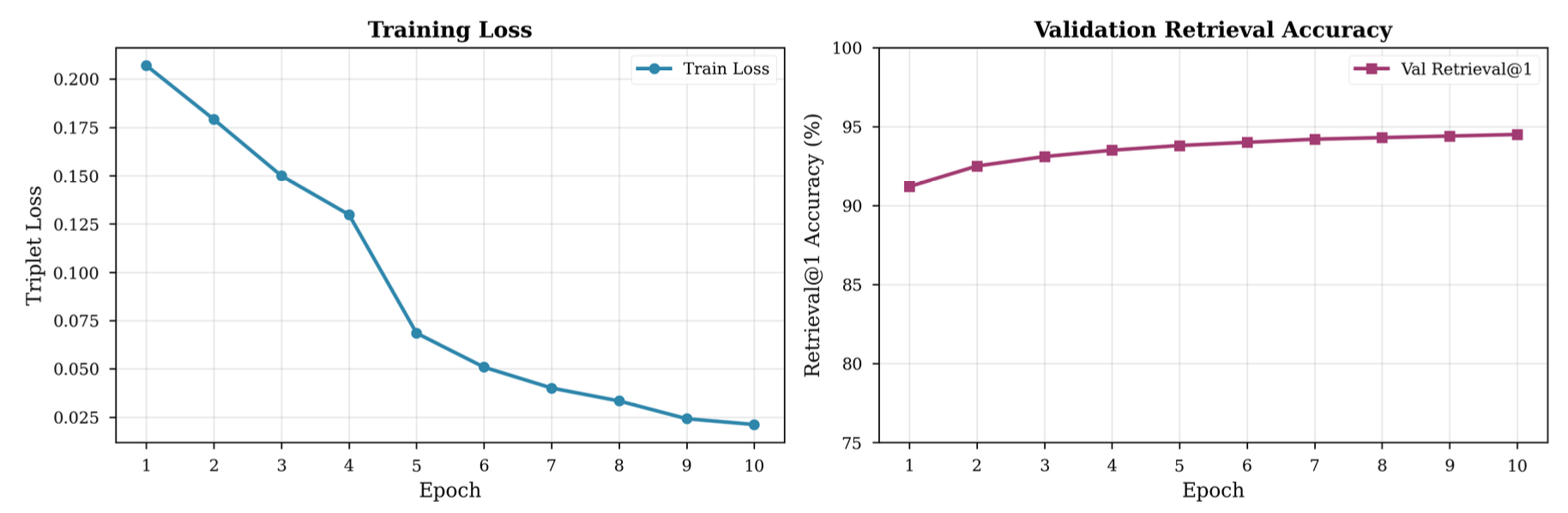}
\caption{Triplet-loss training on MalImg over 10 epochs (Left: training loss.
Right: validation \texttt{retrieval@1} accuracy)}\label{fig:training}
\end{figure}

We train the triplet-loss ResNet-18 encoder for~10 epochs on the MalImg training set,
using a 128-dimensional embedding and the hyperparameters in Table~\ref{tab:train}.
Figure~\ref{fig:training} shows the training progress over the~10 epochs.
The triplet loss (left) decreases steadily over training,
while the validation \texttt{retrieval@1} accuracy (right)
improves from~80.67\%\ in the first epoch to~95.0\%\ at convergence.

The complete test-set results are given in Table~\ref{tab:MI}. We note that
the test \texttt{retrieval@1}
of~94.0\%\ is reported on the~2,537 MalImg test samples 
using a leave-one-out gallery/query protocol, where each test sample is treated as the query
in turn, while the remaining~2,536 test samples form the gallery. This protocol
provides the most comprehensive use of the available test data without contaminating
the training set, and it is consistent with the strict zero-shot evaluation
described in Section~\ref{sect:exp:zeroshot}. 
In an operational deployment, the gallery would instead be
pre-populated with labeled known samples and the queries 
would be new samples.

\begin{table}[!htb]
    \centering
    \caption{MalImg same-domain results (triplet loss, ResNet-18, 128-dim, 10 epochs)}\label{tab:MI}
    \begin{adjustbox}{scale=0.85}
        \begin{tabular}{l|c}
            \toprule
            \multicolumn{1}{c|}{\textbf{Measure}} & \textbf{Result}  \\
            \midrule
            Validation \texttt{retrieval@1} & 95.0\% \\
            Test \texttt{retrieval@1} & 94.0\% \\
            Cluster purity & 0.8265 \\
            Silhouette score & 0.1825 \\
            Separation ratio & 2.8055 \\
            \bottomrule
        \end{tabular}
    \end{adjustbox}
\end{table}

\begin{figure}[!htb]
\centering
\includegraphics[width=0.45\textwidth]{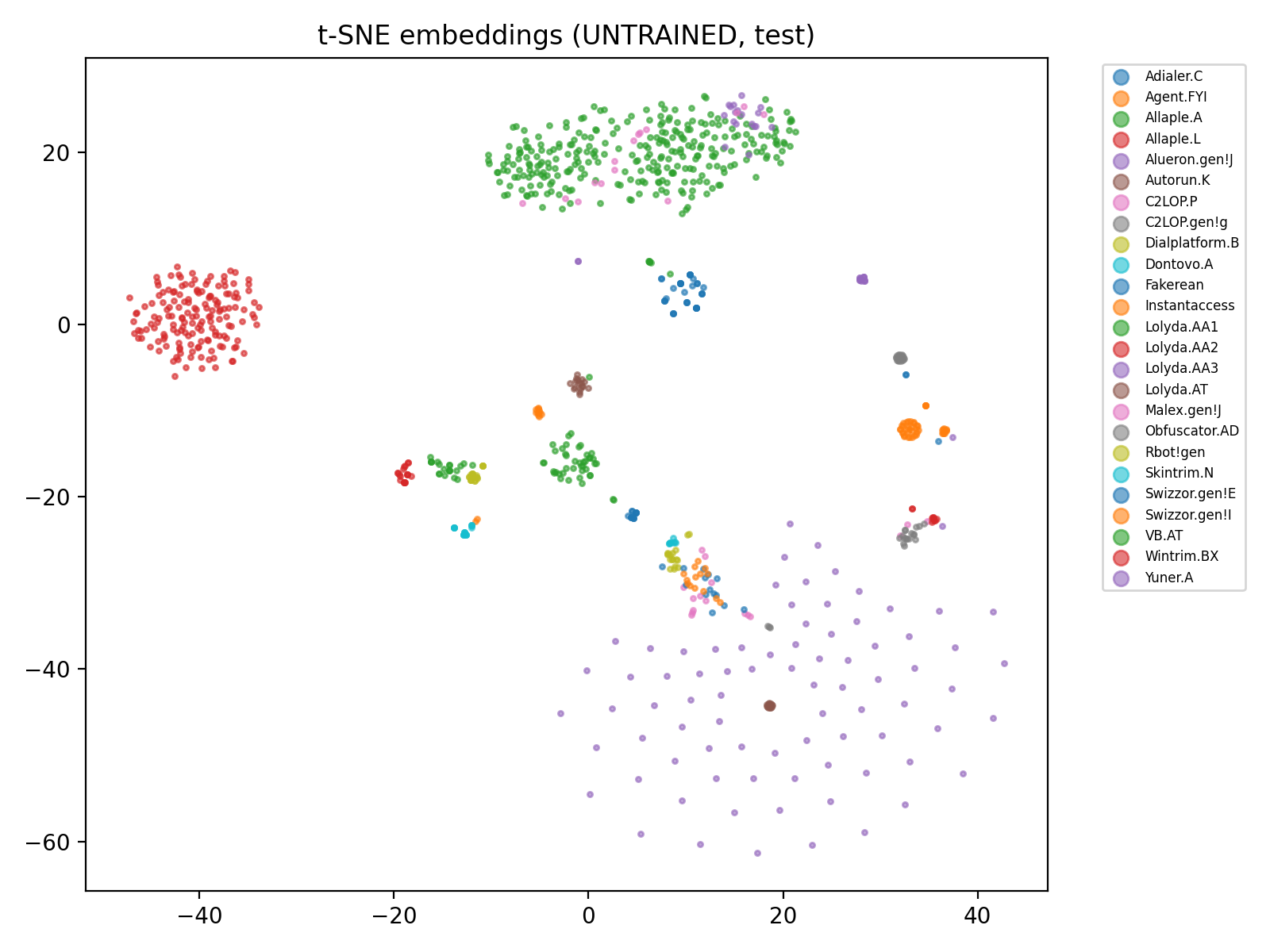}
\includegraphics[width=0.45\textwidth]{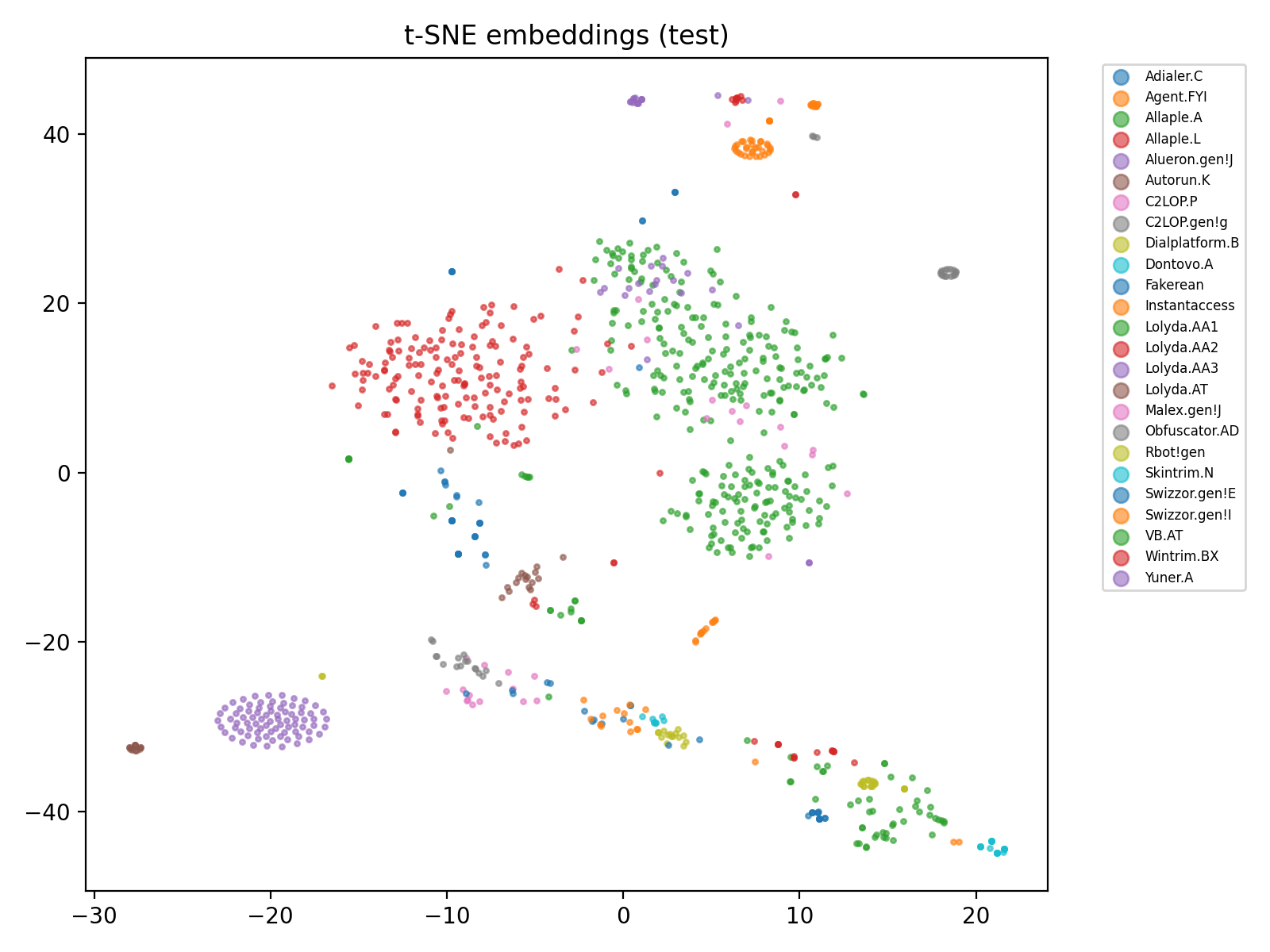}
\caption{t-SNE visualization of embeddings from the untrained encoder (left)
and the trained encoder (right) on the MalImg test set}
\label{fig:tsne}
\end{figure}

The results in Table~\ref{tab:MI} indicate that for the MalImg dataset, the embedding
space shows strong separation between families, with tight intra-class clusters and
clear inter-class boundaries. To visualize the learned space, Figure~\ref{fig:tsne}
shows t-Distributed Stochastic Neighbor Embedding (t-SNE) projections of the
test-set embeddings.
The image on the lefthand side of Figure~\ref{fig:tsne} corresponds to an untrained
encoder, while the image on the righthand side corresponds to the trained encoder.
The trained projection shows visibly tighter, more separated clusters. Since
t-SNE projections can distort high-dimensional structure, the quantitative metrics
in Table~\ref{tab:MI} (cluster purity~0.83, separation ratio~2.81) provide a more
reliable measure of cluster quality than the 2D projection alone.

To understand failure modes on the MalImg dataset, we analyze per-family retrieval accuracy.
Families with the highest accuracy (above~98\%) include \texttt{Allaple.A},
\texttt{Allaple.L}, and \texttt{Yuner.A}, which have relatively large training sets
(2,359, 1,272, and~640 samples respectively) and distinct visual patterns.
Families with lower accuracy (below~85\%) include \texttt{Skintrim.N} (8 test samples),
\texttt{Autorun.K} (12 test samples), and \texttt{Alueron.gen!J} (21 test samples),
indicating that severely limited training data results in poor embeddings.
Class imbalance is thus a significant factor in performance, with rare families
requiring special handling such as class-balanced sampling, data augmentation,
or transfer from related families.

Training, inference, and FAISS search timings for the MalImg experiments are
summarized in Table~\ref{tab:time}. As expected, training is the most time-consuming
step; inference and FAISS search are sub-millisecond per query, indicating that
the technique is practical in a real-world deployment.

\begin{table}[!htb]
    \centering
    \caption{Approximate timings (ResNet-18, MalImg, CPU unless noted)}\label{tab:time}
    \begin{adjustbox}{scale=0.85}
        \begin{tabular}{l|c}
            \toprule
            \multicolumn{1}{c|}{\textbf{Step}} & \textbf{Approximate timing}  \\
            \midrule
            Same-domain MalImg  & \multirow{2}{*}{2--3 hours} \\ 
            (ResNet-18, 10 epochs, CPU) &  \\ \midrule
            Zero-shot & \multirow{3}{*}{$\sim$49 minutes} \\ 
            (small CNN, 80 epochs, Colab GPU) & \\ 
            best checkpoint &  \\ \midrule
            Zero-shot & \multirow{3}{*}{$\sim$3 hours 38 minutes} \\ 
            (ResNet-18, 80 epochs, Colab GPU) & \\ 
            best checkpoint & \\ \midrule
             \multirow{2}{*}{Inference} & 10 ms per image on CPU \\
                            & (1 ms on GPU) \\ \midrule
            \multirow{2}{*}{FAISS search} & less than 1 ms for 10K samples \\
                                   &  (less than 10 ms for 1M samples) \\
            \bottomrule
        \end{tabular}
    \end{adjustbox}
\end{table}

\subsection{Same-Domain Baseline on MalNet}\label{sect:exp:malnet}

For comparison, we also report a same-domain baseline on the larger MalNet-Images-Tiny
dataset. MalNet contains roughly eight times more training samples than MalImg
($\sim$61K versus~$\sim$5.7K) and has~47 families with diverse naming conventions,
making it a substantially harder same-domain target.
We train the triplet-loss ResNet-18 encoder for~30 epochs on the MalNet training set
(61{,}201 images, 47 families), obtaining a validation \texttt{retrieval@1} of~71.8\%.
As a reference point for the rate of convergence, a short 3-epoch run reaches a
validation \texttt{retrieval@1} of~53.69\%\ at a training loss of~0.2002, indicating
that the bulk of the accuracy gain accrues over the subsequent epochs as the
embedding geometry stabilizes.

The lower same-domain accuracy on MalNet relative to MalImg (71.8\%\ versus~94.0\%)
is consistent with three properties of the MalNet dataset: a larger number of families
(47 versus~25), noisier labeling arising from its multi-tag naming convention, and a
broader diversity of malware types (adware, Trojans, backdoors, and spyware).
Notably, despite this lower in-domain accuracy, the MalNet-trained encoder still produces
embeddings that transfer strongly to MalImg, as we demonstrate next.

\subsection{Cross-Domain Generalization}\label{sect:exp:crossdomain}

In real-world malware analysis, models are often trained on one dataset,
but must perform on samples from different sources, time periods,
or collection methodologies. Such a domain shift is a fundamental challenge
in machine learning, and it is particularly important with respect to malware.

To evaluate the generalizability of our approach in a standard cross-domain
setting, we take the MalNet-trained triplet-loss encoder from
Section~\ref{sect:exp:malnet} (the 30-epoch model) and evaluate it
without any retraining on the MalImg test set.
This is a challenging scenario, since the malware families differ substantially
(MalNet has~47 families, MalImg has~25), the two datasets were collected at
different times using different methodologies, and the class distributions
differ. The model must therefore rely on visual patterns that are learnable
from MalNet but also predictive on MalImg, rather than on memorization.

Note that we chose this direction (MalNet $\rightarrow$ MalImg) because MalNet
is the larger dataset and provides a stronger feature learner. The opposite
direction (MalImg $\rightarrow$ MalNet) would be even more
challenging---not only are there fewer training samples in MalImg, but the model must
generalize from~25 training families to~47 test families, doubling the number
of unseen families. In effect, the MalImg $\rightarrow$ MalNet direction is closer
to the strict zero-shot setting that we examine in
Section~\ref{sect:exp:zeroshot}.

The results of this cross-domain experiment are summarized in
Table~\ref{tab:CD}. The \texttt{retrieval@1} accuracy of~88.5\%\ on MalImg demonstrates
that the embeddings learned from MalNet capture visual patterns also present in
the MalImg dataset. While not as strong as the same-domain MalImg result of~94.0\%,
this cross-domain performance is impressive given the dataset differences,
and it is achieved without any retraining or domain adaptation.

\begin{table}[!htb]
    \centering
    \caption{Cross-domain results (train on MalNet, test on MalImg)}\label{tab:CD}
    \begin{adjustbox}{scale=0.85}
        \begin{tabular}{l|c}
            \toprule
            \multicolumn{1}{c|}{\textbf{Measure}} & \textbf{Result}  \\
            \midrule
            Test \texttt{retrieval@1} on MalImg & 88.5\% \\
            Cluster purity & 0.7994 \\
            Silhouette score & 0.2325 \\
            Separation ratio & 1.7968 \\
            Intra-class distance (mean) & 0.0081 \\
            Inter-class distance (mean) & 0.0145 \\
            \bottomrule
        \end{tabular}
    \end{adjustbox}
\end{table}

These strong cross-domain results are likely due to the following factors.
First, the conversion methodology (byte-to-image mapping)
preserves many of the fundamental structural patterns in PE files
(e.g., headers, section boundaries, code regions)
that are universal across malware families, regardless of dataset origin.
Second, the L2 normalization constrains all embeddings to the unit hypersphere,
removing dataset-specific scale differences, and thus the model learns relative
patterns (e.g., ``family A has denser code sections than family B'').
Third, unlike classification models that learn dataset-specific decision boundaries,
triplet loss optimizes for similarity relationships, and the constraint
that samples from the same family are closer than those from different families is universal,
enabling the learned metric to transfer across datasets. Finally,
the separation ratio of nearly~1.80 indicates well-separated clusters,
providing sufficient margin for correct retrieval even for cross-domain embeddings.

\subsection{Strict Zero-Shot Evaluation on Held-Out Families}\label{sect:exp:zeroshot}

This section presents the central experiment of the chapter: zero-shot
recognition of malware families that are entirely absent from the training
data. The distinction from the cross-domain experiment in the previous
section is significant. In the cross-domain case, the train set (MalNet,
47 families) and the test set (MalImg, 25 families) come from different
datasets, but the families themselves overlap in spirit---both are
Windows malware corpora curated using similar methodologies, and many
MalImg family names (e.g., \texttt{Allaple}, \texttt{Yuner}) describe lineages that also
appear under different labels in MalNet, so the encoder can in principle
re-use family-level visual patterns it saw during training. In the
strict zero-shot case, by contrast, the test families
(\Agensla, \Androm, \DCRat, \ldots, \Zenpak) come from an entirely
different corpus (RawMal-TF) and were generated using a different
image-construction procedure (see Section~\ref{sect:data}); none of
them overlap with the~453 training families. The encoder therefore
cannot fall back on having seen these families; it must place them
purely by visual analogy to families it learned from.

\subsubsection{Setup}
Both metric learning models compared here are trained on MalNet and MalImg combined
(453 families across the two datasets after merging, 96{,}769 images) for~80 epochs,
using multi-proxy anchor loss (the refinement described in
Section~\ref{sect:N_finger}) with~16 classes and~4 samples per batch
(P$\times$K sampling), a~256-dimensional embedding, and~4 proxies per class.
The test set is the~17-family grayscale dataset (17{,}000 images), which contains
no families seen during training. We confirmed zero data leakage; there are
no duplicate image hashes between the training and test sets.
The evaluation method uses the same leave-one-out gallery/query protocol as
Section~\ref{sect:exp:samedomain}, where each test sample serves as the
query while the remaining test samples form the gallery.

\begin{figure}[!htb]
\centering
\includegraphics[width=0.7\linewidth]{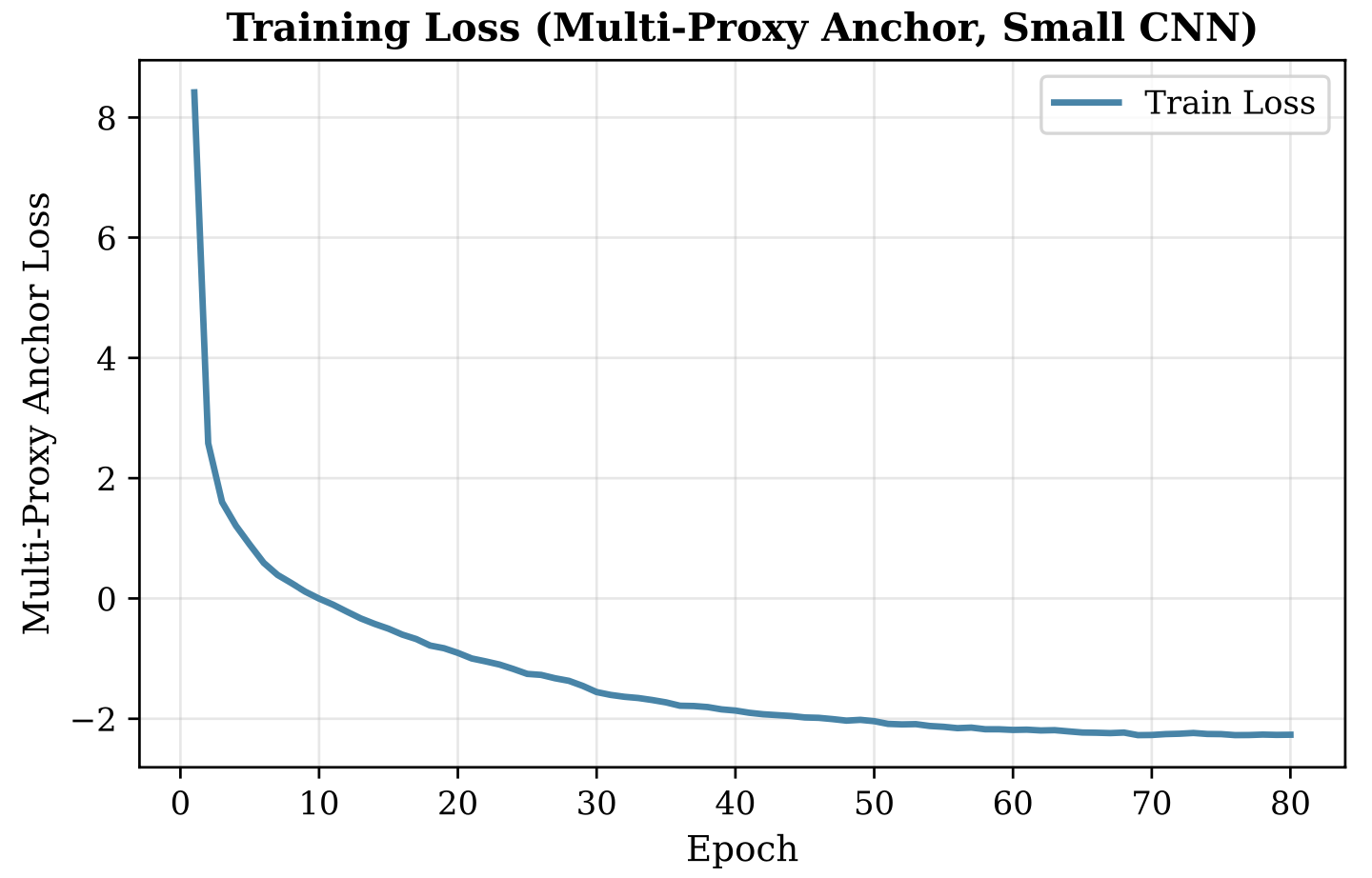}
\caption{Multi-proxy anchor loss over 80 epochs
(small CNN, MalNet+MalImg combined)}\label{fig:training_zs}
\end{figure}

Figure~\ref{fig:training_zs} shows the training loss over~80 epochs for the
small CNN model on the combined training set. The multi-proxy anchor loss decreases
steadily throughout training. After training, the encoder achieves a train-set
\texttt{retrieval@1} of~0.880.

\subsubsection{Results}
The zero-shot test results for both models are given in Table~\ref{tab:zeroshot}.
The random \texttt{retrieval@1} baseline is~5.88\%\ (1/17 families), so both learned
models substantially outperform chance.

\begin{table}[!htb]
    \centering \def\pp{\phantom{$-$}}
    \caption{Zero-shot cross-domain results (train: MalNet+MalImg combined for 80 epochs, test: 17-family grayscale)}\label{tab:zeroshot}
    \begin{adjustbox}{scale=0.85}
        \begin{tabular}{l|cc}
            \toprule
            \multicolumn{1}{c|}{\multirow{2}{*}{\textbf{Measure\hspace*{0.625in}}}} 
            	& \multirow{2}{*}{\textbf{ResNet-18\ }} & \textbf{\ Small CNN}  \\
            		 & & \textbf{(proposed)}  \\
            \midrule
            Train \texttt{retrieval@1} & 93.0\% & 88.0\% \\
            Test \texttt{retrieval@1}  & 70.5\% & \textbf{73.1\%} \\
            Test \texttt{retrieval@3}  & 78.9\% & 81.3\% \\
            Test \texttt{retrieval@5}  & 81.9\% & 85.0\% \\
            Test \texttt{retrieval@10} & 87.0\% & 89.5\% \\
            Open-set AUROC             & \textbf{96.5\%} & 90.5\% \\
            5-way 1-shot accuracy      & 27.7\% & \textbf{33.6\%} \\
            5-way 5-shot accuracy      & 38.7\% & \textbf{45.2\%} \\ \midrule
            Cluster purity             & \pp0.267\ \ \  & \textbf{\pp0.309\ \ \ } \\
            Silhouette score           & $-$0.070\ \ \ & \textbf{$-$0.066\ \ \ } \\
            Separation ratio           & \pp1.082\ \ \ & \textbf{\pp1.118\ \ \ } \\ \midrule
            Random \texttt{retrieval@1} baseline & \multicolumn{2}{c}{5.88\%} \\
            \bottomrule
        \end{tabular}
    \end{adjustbox}
\end{table}

Both backbones yield low intrinsic clustering scores on the zero-shot test set,
which is expected given the difficulty of the strict zero-shot setting, in which the
test families were never observed during training. The small CNN is slightly better
than ResNet-18 on all three intrinsic metrics (cluster purity~0.309 versus~0.267,
silhouette score~$-0.066$ versus~$-0.070$, and separation ratio~1.118 versus~1.082),
consistent with its higher retrieval accuracy. In both cases the separation ratio
exceeds~1, confirming that, on average, inter-family distances are larger than
intra-family distances even for these unseen families, while the near-zero silhouette
scores indicate that the family clusters, though present, overlap substantially in
the embedding space.

Several observations are worth highlighting. First, the lightweight small CNN
outperforms ResNet-18 on test \texttt{retrieval@1} (73.1\%\ vs 70.5\%) and on
both few-shot metrics, despite having lower training \texttt{retrieval@1}
(88.0\%\ vs 93.0\%). This pattern---higher training accuracy but lower test
accuracy for the deeper model---suggests that ResNet-18 overfits more to features
specific to the training families, and that a smaller backbone is forced to
learn more generalizable features. We conclude that for the strict zero-shot
setting on grayscale malware images, model capacity is not the primary bottleneck.
Second, ResNet-18 achieves a higher open-set AUROC (96.5\%\ vs 90.5\%), meaning
it is better at flagging samples that do not belong to any gallery family.
The two models thus offer complementary strengths---the small CNN is the
better retriever, while ResNet-18 is the better novelty detector. A
natural follow-up would be to combine these models in a two-stage pipeline.
That is, we could have the ResNet-18 model
act as an open-set gate, rejecting queries whose nearest-gallery
similarity falls below a threshold (i.e., flagging them as belonging to
an unknown family), with the small CNN then performing the retrieval for
queries that pass the gate. We leave a controlled evaluation of such an
ensemble to future work.

A practical note on convergence is also worth highlighting. The small CNN reaches
its best validation \texttt{retrieval@1} after only $\sim$49 minutes of GPU
training and does not improve further over the remaining $\sim$2 hours
42 minutes of the 80-epoch budget, whereas ResNet-18 continues to improve
until near the end of its $\sim$3 hour 38 minute run. The lightweight
backbone is therefore not only the more accurate zero-shot retriever
on this benchmark but also roughly~$4\times$ cheaper to train to its
best checkpoint.

Compared to the same-domain MalImg result (94\%) and the cross-domain MalNet $\rightarrow$
MalImg result (88.5\%), the strict zero-shot result of~73.1\%\ shows the expected
drop in accuracy as the evaluation becomes progressively harder. The drop is
substantial but not catastrophic, and the absolute level (more than~12$\times$ the
random baseline) confirms that the learned embedding space does capture
generalizable visual structure.

\subsubsection{Confusion analysis}
\begin{figure}[!htb]
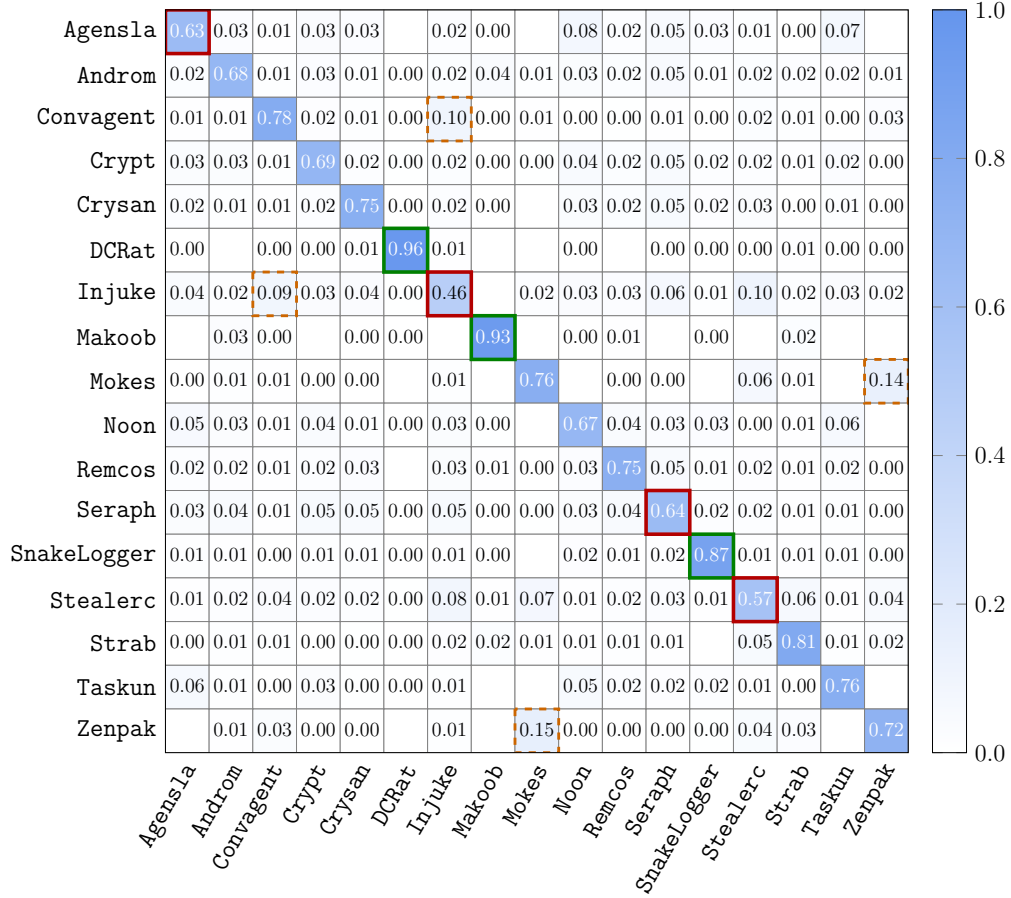

\centering
\begin{adjustbox}{scale=1.0}
\input figures/conf_MalImg.tex
\end{adjustbox}
\caption{Retrieval confusion matrix for the~17-family grayscale dataset, zero-shot.
Rows are actual families; columns are predicted families.
Each cell shows the proportion of queries retrieved at rank~1.
\textcolor{green!50!black}{\textbf{Green}} borders indicate the highest-accuracy families,
\textcolor{red!70!black}{\textbf{red}} borders the lowest,
and \textcolor{orange!80!black}{\textbf{orange}} dashed borders highlight
the most significant cross-family confusions.}\label{fig:confusion}
\end{figure}

Figure~\ref{fig:confusion} provides a retrieval confusion matrix for the~17-family
grayscale test set. Each row sums to~1.00 and the entries on the main diagonal show the
per-family \texttt{retrieval@1} accuracy. Careful analysis reveals that confusion
occurs primarily between visually similar families (e.g., \Zenpak\ and \Mokes,
and \Convagent\ and \Injuke) that share structural byte patterns when rendered as
grayscale images.

Families with the highest accuracy are \DCRat\ (96.4\%), \Makoob\
(93.3\%), and \SnakeLogger\ (86.9\%), which have visually distinct byte
patterns. The lowest-accuracy families are \Injuke\ (45.5\%),
\Stealerc\ (57.0\%), and \Agensla\ (63.4\%), with \Seraph\ (64.0\%) also
yielding poor results. These families distribute a substantial fraction of their queries
across several visually similar families. In particular, \Injuke\ and
\Convagent\ mutually confuse each other (cross-retrieval rates of~0.088
and~0.095), as do \Mokes\ and \Zenpak\ (0.139 and~0.151),
suggesting that these family pairs share similar low-level structural
patterns in their grayscale representations.
This analysis confirms that visual similarity between families is the primary
driver of retrieval errors, motivating future work on fine-grained
discriminative training objectives.

\subsection{Robustness Analysis}

We evaluate robustness of our trained models under the image perturbations listed in Table~\ref{tab:pert}.
Intuitively, we expect that rotation will be the most challenging perturbation for our model,
while small amounts of noise should have minimal impact.

\begin{table}[!htb]
    \centering
    \caption{Perturbations for robustness analysis}\label{tab:pert}
    \begin{adjustbox}{scale=0.85}
        \begin{tabular}{l|l}
            \toprule
            \textbf{Perturbation} & \multicolumn{1}{c}{\textbf{Parameters}}  \\
            \midrule
            Gaussian noise & Standard deviation 0.1 added independently to each pixel \\ \midrule
            \multirow{2}{*}{Random erasing} & Randomly zeroes a rectangle;  \\
            	& probability~0.1, area scale~$(0.02, 0.2)$ \\ \midrule
            Rotation & Image rotated by a random angle in~$[-10^{\circ}, +10^{\circ}]$ \\ \midrule
            \multirow{2}{*}{Crop} & Random resized crop then resize back to $224\times224$;  \\
            	& area scale~$(0.7, 1.0)$ \\
             \bottomrule
        \end{tabular}
    \end{adjustbox}
\end{table}

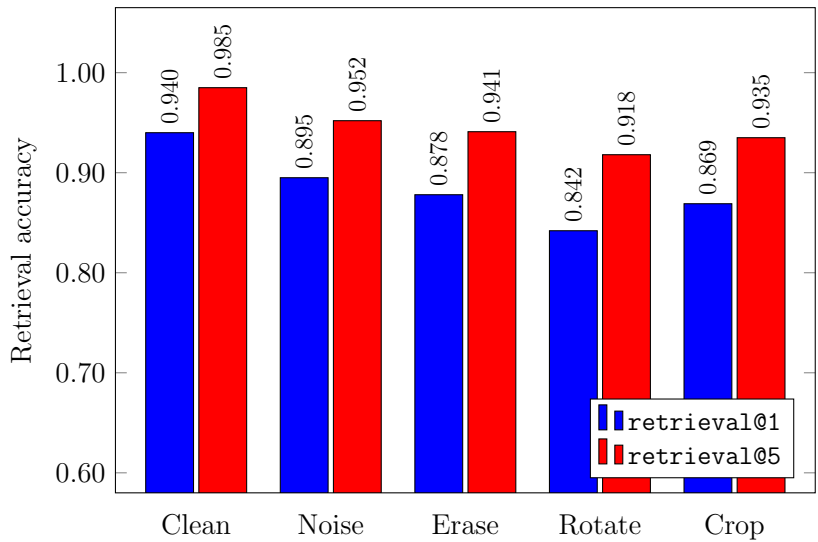
\begin{figure}[!htb]
\centering
\begin{tikzpicture}[scale=1.0, every node/.style={scale=1.0}]
\pgfkeys{/pgf/number format/.cd,1000 sep={}}
\begin{axis}[
        width  = 0.7*\textwidth,
        height = 8.0cm,
        ymin=0.58,ymax=1.065,
        ytick={0.60, 0.70, 0.80, 0.90, 1.00},
        major x tick style = transparent,
        ybar=5*\pgflinewidth,
        bar width=18.0pt,
        ylabel = {Retrieval accuracy},
        ylabel style = {scale = 0.9},
        symbolic x coords={A, B, C, D, E},
        xticklabels={Clean, Noise, Erase, Rotate, Crop},
	y tick label style={scale=0.9,
    		/pgf/number format/.cd,
   		fixed,
   		fixed zerofill,
    		precision=2},
        xtick = data,
        x tick label style={scale=0.9,
		},
        nodes near coords,
        every node near coord/.append style={rotate=90, scale=0.785,
        								   anchor=west, 
								   /pgf/number format/.cd,
								   fixed,
								   fixed zerofill,
								   precision=3},
        enlarge x limits=0.15,
        legend cell align=left,
        legend pos=south east,
        legend style={nodes={scale=0.85},
        },
]
\addplot [fill=blue,opacity=1.00]
coordinates {
(A, 0.940)
(B, 0.895)
(C, 0.878)
(D, 0.842)
(E, 0.869)
};
\addlegendentry{\texttt{retrieval@1}}
\addplot [fill=red,opacity=1.00]
coordinates {
(A, 0.985)
(B, 0.952)
(C, 0.941)
(D, 0.918)
(E, 0.935)
};
\addlegendentry{\texttt{retrieval@5}}
\end{axis}
\end{tikzpicture}
\caption{Robustness under image perturbations (same-domain MalImg)}\label{fig:robust}
\end{figure}

For each perturbation in Table~\ref{tab:pert},
we perturb the test samples as indicated and record the retrieval accuracy.
Figure~\ref{fig:robust} compares the \texttt{retrieval@1} and \texttt{retrieval@5}
accuracies thus obtained on the same-domain MalImg test set.
As expected, rotation is the most challenging perturbation, as it fundamentally
alters the spatial arrangement of byte patterns in the image. We note that
\texttt{retrieval@1} accuracy remains above~84\%\ in all cases (compared with~94\%\
on the clean, i.e., unperturbed, test data), and \texttt{retrieval@5} accuracy remains above~91\%\
(compared with~98.5\%\ for the clean data).
Gaussian noise and random erasing have a smaller impact since the metric learner
relies on global structural patterns rather than pixel-level precision.
These results indicate that the model achieves a reasonable level of robustness
to common image perturbations.

\subsection{Comparison of Three Approaches}

Recall that for the SVM, 256 PCA features are obtained by flattening the~$224\times224$
grayscale image and reducing to~256 dimensions; for the deep-learning baseline, the ResNet-18 backbone
yields a~512-dimensional feature vector to which we apply a softmax output layer
corresponding to the~25 classes in MalImg; and for metric learning, we project the
backbone features to~128-dimensional L2-normalized embeddings via the MLP head.
The model parameters are summarized in Table~\ref{tab:approaches};
see Section~\ref{sect:meth} for more details.

\begin{table}[!htb]
    \centering
    \caption{Summary of model parameters}\label{tab:approaches}
    \begin{adjustbox}{scale=0.85}
        \begin{tabular}{c|ll}
            \toprule
            \textbf{Technique} & \textbf{Parameter} & \textbf{Details} \\ \midrule
            \multirow{5}{*}{Classic ML}
            & Method & SVM \\
            & Kernel & RBF \\
            & Features & 256 (after PCA from 50{,}176) \\
            & Preprocessing & \texttt{StandardScaler} normalization \\
            & Training & supervised classification, one-vs-rest \\ \midrule
            \multirow{4}{*}{Deep learning}
            & Method & ResNet-18 (supervised classification head) \\
            & Architecture & ResNet-18 backbone, 25-class linear head \\
            & Loss & cross-entropy loss \\
            & Training & supervised learning \\ \midrule
            \multirow{5}{*}{Metric learning}
            & Method & ResNet-18 / small CNN encoder \\
            & Architecture & ResNet-18 or lightweight CNN backbone \\
            & Loss & batch-hard triplet loss $\rightarrow$ multi-proxy anchor loss \\
            & Embedding dim & 128 (same/cross-domain), 256 (zero-shot) \\
            & Training & metric learning for similarity optimization \\
            \bottomrule
        \end{tabular}
    \end{adjustbox}
\end{table}

Table~\ref{tab:compare} provides a detailed comparison of the results
for the three approaches considered in this chapter.
The headline metric for each paradigm is the strongest
result it can naturally produce: same-domain MalImg \texttt{retrieval@1} for the
SVM and the ResNet-18 classifier (requiring fixed, known classes),
and the same-domain, cross-domain, and zero-shot \texttt{retrieval@1} values for
metric learning. The same-domain numbers (94.0\%, 91.0\%, 86.0\%)
are directly comparable; the cross-domain (88.5\%) and zero-shot (73.1\%) numbers
cannot be produced by the SVM or the ResNet-18 classifier without retraining, and are
therefore unique capabilities of the metric-learning approach.

\begin{table}[!htb]
    \centering
    \caption{Comparison of the three approaches and their capabilities
    (``n/a'' indicates a capability the paradigm cannot provide without retraining}\label{tab:compare}
    \begin{adjustbox}{scale=0.85}
        \begin{tabular}{l|ccc}
            \toprule
            \multirow{2}{*}{\textbf{Description}}
            	& \multirow{2}{*}{\textbf{Classic ML}} & \textbf{Deep} & \textbf{Metric} \\ 
            	& & \textbf{Learning} & \textbf{Learning} \\ \midrule
            Same-domain & \multirow{3}{*}{86.0\%} & \multirow{3}{*}{91.0\%} & \multirow{3}{*}{94.0\%} \\
            	MalImg & \\
            	\texttt{retrieval@1} & \\ \midrule
            Cross-domain & \multirow{3}{*}{n/a} & \multirow{3}{*}{n/a} & \multirow{3}{*}{88.5\%} \\
            	MalNet$\to$MalImg & \\
            	\texttt{retrieval@1} & \\ \midrule
            Zero-shot 17-family & \multirow{2}{*}{n/a} & \multirow{2}{*}{n/a} & \multirow{2}{*}{73.1\%} \\
            	\texttt{retrieval@1} & \\ \midrule
            Training time & \multirow{2}{*}{$<$0.1\,h} & \multirow{2}{*}{$\sim$3\,h 38\,min} & $\sim$49\,min \\
            80 epochs, $1\times\mbox{Colab GPU}$ & & & (small CNN) \\ \midrule
            Inference time (ms, GPU) & 5 & 10 & 1 \\ \midrule
            Memory usage (MB) & 50 & 150 & 150 \\ \midrule
            Zero-shot support & No & No & Yes \\
            Similarity score & No & No & Yes \\
            Interpretability & Low & Low & High \\
            Retrain for new families & Yes & Yes & No \\ \midrule
            Strength & Fast, simple & Deep features & Flexible, scalable \\
            Weakness & Fixed classes & Fixed classes & Requires gallery \\
            \bottomrule
        \end{tabular}
    \end{adjustbox}
\end{table}

Based on these results, we observe that the classic ML approach using SVM may be
advantageous when the dataset is small, the classes are fixed and known in advance,
and fast training is critical. A deep learning approach based on ResNet-18 with
a softmax head may be preferred when maximum accuracy is needed in a closed-set
setting. Our metric learning approach is preferable when new malware families
appear frequently (zero-shot requirement), similarity scores and interpretability
are important, the gallery contains a large number of samples (FAISS scalability),
or multiple downstream tasks must be considered (e.g., retrieval, clustering, visualization).

\begin{figure}[!htb]
\centering
\begin{tikzpicture}[scale=1.0, every node/.style={scale=1.0}]
\pgfkeys{/pgf/number format/.cd,1000 sep={}}
\begin{axis}[
	axis y line*=left,
        width  = 0.6*\textwidth,
        height = 8.0cm,
        ymin=0.58,ymax=1.035,
        ytick={0.60, 0.70, 0.80, 0.90, 1.00},
        major x tick style = transparent,
        ybar=5*\pgflinewidth,
        bar width=21.5pt,
        bar shift=-12.0pt,
        ylabel = {Test accuracy},
        ylabel style = {scale = 0.9},
        symbolic x coords={A, B, C},
        xticklabels={Classic ML (SVM), DL (ResNet-18), Metric learning},
	y tick label style={scale=0.9,
    		/pgf/number format/.cd,
   		fixed,
   		fixed zerofill,
    		precision=2},
        xtick = data,
        x tick label style={scale=0.9,        	
        		rotate=60,
		anchor=north east,
		inner sep=0mm
		},
        nodes near coords,
        every node near coord/.append style={rotate=90, scale=0.785,
        								   anchor=west, 
								   /pgf/number format/.cd,
								   fixed,
								   fixed zerofill,
								   precision=3},
        enlarge x limits=0.265,
        legend cell align=left,
        legend pos=south east,
        legend style={nodes={scale=0.785},
        },
]
\addplot [fill=blue,opacity=1.00]
coordinates {
(A, 0.860)
(B, 0.910)
(C, 0.731)
}; \label{AAA}
\end{axis}
\begin{axis}[
	axis y line*=right,
        width  = 0.6*\textwidth,
        height = 8.0cm,
        ymin=0.0,ymax=4.0,
        ytick={0.0, 1.0, 2.0, 3.0, 4.0},
        major x tick style = transparent,
        ybar=5*\pgflinewidth,
        bar width=21.5pt,
        bar shift=12.0pt,
        ylabel = {Training time (hours)},
        ylabel style = {scale = 0.9},
        symbolic x coords={A, B, C},
	xticklabels=empty,
	y tick label style={scale=0.9,
    		/pgf/number format/.cd,
   		fixed,
   		fixed zerofill,
    		precision=1},
        nodes near coords,
        every node near coord/.append style={rotate=90, scale=0.785,
        								   anchor=west, 
								   /pgf/number format/.cd,
								   fixed,
								   fixed zerofill,
								   precision=1},
        enlarge x limits=0.265,
        legend cell align=left,
        legend pos=south east,
        legend style={nodes={scale=0.785},
        },
]
\addlegendimage{/pgfplots/refstyle=AAA}
\addlegendentry{Test accuracy}
\addplot [fill=red,opacity=1.00]
coordinates {
(A, 0.1)
(B, 2.3)
(C, 3.5)
};
\addlegendentry{Training time}
\end{axis}
\end{tikzpicture}
\caption{Comparison of three machine learning approaches}
\label{fig:three_methods}
\end{figure}
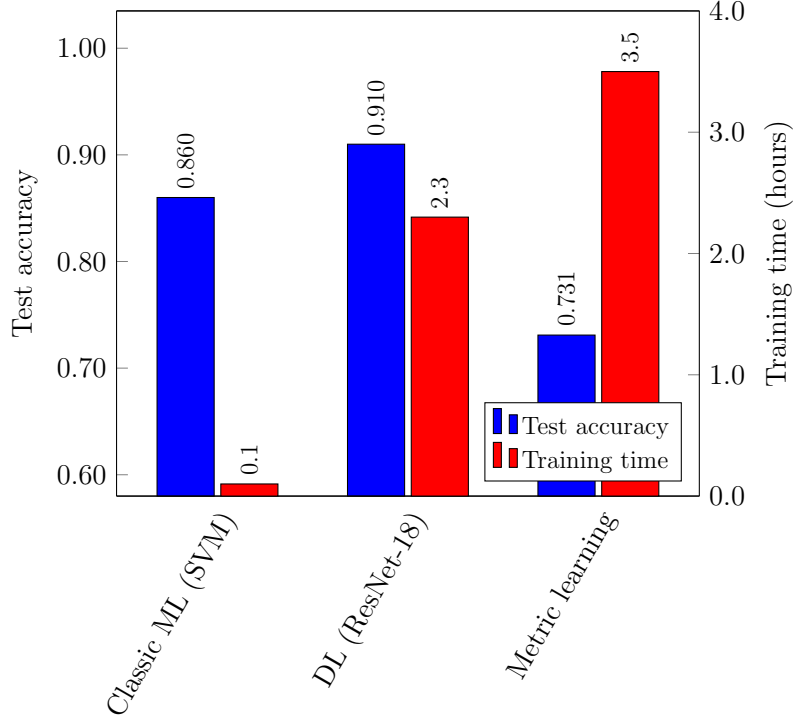

Figure~\ref{fig:three_methods} provides a comparison of the three approaches in the form of a bar graph.
For malware classification in a dynamic threat environment, where new malware families
emerge, these results clearly show that metric learning provides
a practical solution---it matches or exceeds the classification approaches in
the same-domain setting while remaining usable in cross-domain and zero-shot settings
where the classification approaches do not apply.

\subsection{Ablation Studies}\label{sect:ablation}

In this section, we discuss key design choices for our metric-learning models,
covering both backbone selection
and same-domain hyperparameter sensitivity.

\subsubsection{Backbone Choice}

In the same-domain MalImg setting, comparing ResNet-18 to the deeper ResNet-50,
we find that both achieve a test \texttt{retrieval@1} of approximately~94.0\%, but
ResNet-50 requires~4.1 hours of CPU training time versus~2.3 hours for ResNet-18---a
roughly~78\%\ increase with no accuracy improvement. We conclude that ResNet-18 is
sufficient for MalImg.

In the strict zero-shot setting, we compare ResNet-18 to a lightweight small CNN.
Both models are trained on the combined MalNet+MalImg training pool
(for~80 epochs with multi-proxy anchor loss and a~256-dimensional
embedding), then evaluated on the held-out 17-family grayscale dataset.
The results are summarized in Table~\ref{tab:resEmbed}.

\begin{table}[!htb]
    \centering \def\pp{\phantom{$-$}}
    \caption{Backbone comparison in the strict zero-shot setting (80 epochs,
    multi-proxy anchor loss, 256-dim embedding, evaluated on 17-family
    grayscale)}\label{tab:resEmbed}
    \begin{adjustbox}{scale=0.85}
        \begin{tabular}{lcc}
            \toprule
            \textbf{Config} & \textbf{ResNet-18} & \textbf{Small CNN (proposed)} \\ \midrule
            Backbone & ResNet-18 & lightweight CNN \\
            Embedding dim & 256 & 256 \\
            Loss & multi-proxy anchor & multi-proxy anchor \\
            Epochs & 80 & 80 \\
            Batch (P$\times$K) & $16\times4$ & $16\times4$ \\ \midrule
            Train \texttt{retrieval@1} & 93.0\% & 88.0\% \\
            Test \texttt{retrieval@1} & 70.5\% & \textbf{73.1\%} \\
            Open-set AUROC & \textbf{96.5\%} & 90.5\% \\
            5-way 1-shot & 27.7\% & \textbf{33.6\%} \\
            5-way 5-shot & 38.7\% & \textbf{45.2\%} \\
            Cluster purity & \pp0.267\pp & \textbf{\pp0.309\pp} \\
            Silhouette score & $-$0.070\pp & \textbf{$-$0.066\pp} \\
            Separation ratio & \pp1.082\pp & \textbf{\pp1.118\pp} \\
            Training to best checkpoint & 3\,h 38\,min & \textbf{49\,min} \\
            \bottomrule
        \end{tabular}
    \end{adjustbox}
\end{table}

The small CNN outperforms ResNet-18 on test \texttt{retrieval@1}
and on both few-shot metrics, despite a higher train/test gap.
This suggests that ResNet-18 overfits more to features specific to the
training families, and that a smaller
backbone generalizes better in this zero-shot cross-domain setting.
ResNet-18 achieves higher open-set AUROC.
Here, open-set AUROC measures
the ability of the nearest-gallery similarity score to separate
in-gallery queries (samples from a family present in the gallery) from
out-of-gallery queries (samples from a family that is not). A higher
AUROC therefore means that the score distributions for these two
populations are better separated under ResNet-18 than under the small
CNN, so a threshold on the similarity score is a more reliable indicator
of ``does this sample belong to a known family?''
The choice of backbone therefore depends on the deployment scenario.
The small CNN is preferred when the primary goal is retrieving the
correct family among samples that all do belong to the gallery, while
ResNet-18 is preferred when the detector must also flag samples that
belong to no known family.

\subsubsection{Same-Domain Hyperparameter Sensitivity}

We additionally ran a same-domain hyperparameter sweep on MalImg with triplet loss
so as to characterize sensitivity to the embedding dimension, the triplet margin, and the
batch size. We vary each hyperparameter individually while holding the others fixed
at their default values. The results of these experiments
are reported in Table~\ref{tab:hyper_ablation}. As shown in the rightmost column,
the cluster-purity trend for the embedding-dimension sweep mirrors the
\texttt{retrieval@1} trend, with larger dimensions yielding slightly purer
clusters. We did not log per-run cluster purity for the margin and batch-size
sweeps, so those entries are marked ``---''.

\begin{table}[!htb]
    \centering
    \caption{Same-domain (MalImg) hyperparameter ablation (triplet loss, ResNet-18, 10 epochs)}\label{tab:hyper_ablation}
    \begin{adjustbox}{scale=0.85}
        \begin{tabular}{c|ccc}
            \toprule
            \multirow{2}{*}{\textbf{Hyperparameter}} & \multirow{2}{*}{\textbf{Value}} 
            		& \textbf{Test} & \multirow{2}{*}{\textbf{Cluster purity}} \\
            		& & \textbf{\texttt{\ retrieval@1\ }} & \\ \midrule
            \multirow{3}{*}{Embedding dimension} & \zz64 & 91.2\% & 0.78 \\
            & 128 & 94.0\% & 0.81 \\
            & 256 & 94.1\% & 0.82 \\ \midrule
            \multirow{3}{*}{Margin~$\alpha$} & 0.1 & 92.1\% & --- \\
            & 0.2 & 94.0\% & --- \\
            & 0.5 & 92.8\% & --- \\ \midrule
            \multirow{3}{*}{Batch size} & \zz32 & 91.8\% & --- \\
            & \zz64 & 94.0\% & --- \\
            & 128 & 94.2\% & --- \\
            \bottomrule
        \end{tabular}
    \end{adjustbox}
\end{table}

From the results in Table~\ref{tab:hyper_ablation}, we observe that~128 dimensions
provide an excellent balance, as~64 dimensions lose discriminative capacity
while~256 dimensions offer only marginal improvement at higher computational cost.
The~128-dimensional embedding is consistent with the
face-recognition literature~\cite{schroff2015facenet}, which has converged on
embedding sizes in the~128--512 range for similar efficiency reasons.
For the strict zero-shot setting we adopt~256 dimensions, reflecting the slight
gain in capacity that becomes useful when generalizing to entirely unseen families.

With respect to the margin, we find that~$\alpha=0.2$ provides the best results---a
small margin of~0.1 allows families to overlap, while the larger margin of~0.5
makes the constraint too difficult to satisfy, leading to slower convergence and
suboptimal embeddings. This finding is consistent with values reported in the
metric-learning literature~\cite{hermans2017defense,schroff2015facenet}, where
margins in the range of~0.2 to~0.3 are typical.

For the batch size---in all cases using 128-dimensional embeddings and a margin
of~0.2---we find that~64 provides sufficient diversity for batch-hard mining
(i.e., enough candidates for hard negatives) while maintaining stable gradients.
Batches of size~128 offer only marginal improvement, while requiring substantially
more memory. On the other hand, smaller batches of size~32 provide insufficient
diversity for effective hard negative mining, leading to lower accuracy.

Finally, for the classic-ML baseline we examined the PCA target dimensionality used
prior to the SVM. We find that projecting the standardized~50,176-dimensional pixel vectors
onto~256 principal components retains more than~95\%\ of the cumulative explained variance
on MalImg. Reducing to~64 or~128 components discards discriminative variance and lowers
SVM accuracy. On the other hand, increasing to~512 components provides no measurable accuracy
improvement while increasing both the PCA and SVM fitting times. We therefore fix the
PCA dimensionality at~256 for all SVM experiments.

\subsubsection{Reproducibility and Confidence}

All results in this chapter are derived from single training runs with fixed random
seeds (seed~42 for the zero-shot experiments, and the default \texttt{PyTorch} seed for
same-domain runs). While this ensures bitwise reproducibility, single-run point
estimates cannot capture the run-to-run variance inherent in stochastic
optimization, and we therefore do not attach confidence intervals to the reported
numbers. Repeating all experiments across multiple seeds to report mean and
variance is left to future work. We also note that wall-clock training times
were not systematically logged via \texttt{ended\_at} markers; the timings
reported in Table~\ref{tab:time} and Table~\ref{tab:compare} are derived from
the run-config \texttt{started\_at} timestamps combined with the checkpoint
file modification times, which gives reliable values to within a few minutes.
The consistent ordering of methods
across multiple evaluation metrics (SVM:~86\%, ResNet:~91\%, triplet
metric learning:~94\%\ same-domain; metric learning generalizing to~88.5\%\
cross-domain and~73.1\%\ zero-shot) nonetheless provides confidence in the
qualitative conclusions.


\section{Discussion}\label{sect:diss}

The previous section provides empirical evidence supporting the effectiveness
of the design choices adopted in our metric learning malware classification model.
In this section, we discuss the advantages and limitations of the proposed metric learning approach.

\subsection{Advantages of Metric Learning}

Metric learning is particularly well-suited for malware classification
because it addresses some of the fundamental
limitations of classification-based approaches. For example,
traditional classifiers learn a function~$f: \mathcal{X} \rightarrow \{1, \ldots, C\}$ that maps inputs to
one of~$C$ predefined classes. When a new malware family emerges, the classifier cannot identify
it without retraining, which is computationally expensive and requires labeled data.
In contrast, metric learning learns a function of the form~$f: \mathcal{X} \rightarrow \mathbb{R}^d$ that maps
inputs to a continuous embedding space. This embedding
space merely encodes similarity relationships, rather than determining
class membership, and hence new families can be added without retraining. When a new family appears,
the samples from this new family are encoded as~$z_{\text{new}} = f_\theta(x_{\text{new}})$.
We then search the gallery for the nearest neighbors of these samples.
If the nearest neighbors belong to the same (new) family, we group them together.
Thus, the new family becomes part of the gallery without retraining.
This zero-shot capability is possible because the embedding space captures
universal visual patterns (e.g., code structure, entropy patterns) that generalize across families seen during training.
Thus, even though the embedding space is shaped by the training families,
its structure can be extended to visually similar unseen families,
as demonstrated by our zero-shot results on the~17-family grayscale dataset.

Another weakness of classification models is that they only provide a class label and
confidence score, which offers limited interpretability. Metric learning provides much richer information,
including similarity scores (e.g., ``0.95 similarity to family~$X$''), which provide
more nuanced information than a hard class label.
Nearest neighbors in metric learning show us which samples are most similar,
thus providing explainable evidence for a classification decision.
In metric learning
the distance to the nearest neighbor provides a natural confidence measure---the
closer the neighbor, the higher the confidence.
Ambiguity detection is another inherent aspect of metric learning.
If a query is nearly equidistant from multiple families, we can flag it as ambiguous,
which is valuable information for analysts.

As malware databases grow, classification becomes increasingly complex.
In contrast, metric learning with FAISS provides for sub-linear search,
since approximate nearest-neighbor algorithms
achieve sub-linear~$O(\log N)$ complexity~\cite{johnson2019billion}.
In addition, the gallery is indexed once but can be queried indefinitely without recomputation,
and new samples can be added to the index incrementally without rebuilding the model.

With metric learning, the same embedding space can be used for multiple tasks.
Examples of such tasks include
retrieval (find similar malware samples, which is the primary use case considered in this chapter),
clustering (group samples into meaningful families),
visualization (project embeddings using t-SNE for exploratory analysis),
anomaly detection (identify outliers), and
family evolution tracking (measure how families evolve over time by tracking embedding distances).
This flexibility is not possible with classification models, where each of these tasks
would require a separate model.

Metric learning is grounded in the theory of metric spaces and manifold learning, which
provides a solid theoretical foundation. This is in contrast to deep learning,
where models are often poorly understood, with limited supporting theory.
In our application of metric learning, malware families form manifolds consisting of
low-dimensional structures embedded in a high-dimensional space.
Metric learning learns a mapping that
preserves the manifold structure, in the sense that
samples from the same family remain close under the mapping.
The mapping also separates different manifolds, i.e., samples from different families are relatively far apart,
and the mapping provides a metric that respects these relationships.
This geometric perspective is potentially more informative than a mere classification,
since it captures the continuous nature of malware relationships (i.e., families can be more or less similar),
rather than treating them as discrete categories.

\subsection{Limitations of Metric Learning}\label{sect:lim}

Our metric learning approach has several limitations. Most of these limitations are
related to the fact that malware samples consist of inherently non-image data,
while a few of the limitations are more technical in nature. In this section, we discuss
examples of such limitations, but this is not intended to serve as an exhaustive list.

It is not unusual for malware samples to be packed or encrypted. Images derived
from such samples will lack much of the structure that image-based approaches rely on,
and hence our technique---which relies on byte-level structure---is not suitable for this type of malware.
However, it is important to note that this is a fundamental limitation of any image-based malware analysis
method, and not specific to our metric learning approach.

Metamorphic malware uses code transformation techniques (register swapping, instruction substitution, dead
code insertion, etc.) that preserve functionality but alter the binary representation. While some structural patterns
may persist, carefully-designed metamorphic malware can blur the visual similarity between variants,
making image-based techniques infeasible. Again, this is a general limitation on image-based malware
analysis.

Distinct malware families with very few samples
(e.g., \texttt{Skintrim.N} in our MalImg test set has only~8 samples)
present difficulties. With extremely limited family training data, the embedding space is
unlikely to be well-learned, leading to poor retrieval performance. It is common in malware
datasets to find a number of families with limited samples.

Our cross-domain experiments showed reasonably strong generalization between
MalImg and MalNet, and our strict zero-shot experiments demonstrated meaningful
(though weaker) generalization to entirely unseen families. However, such
generalization will likely be poor on datasets that employ
different binary-to-image conversion methodologies, as well as
for datasets with significantly different types of malware families
(e.g., mobile malware versus desktop malware). A thorough characterization
of zero-shot transfer under these stronger shifts is a promising direction
for follow-up work.

Generally, images are resized to~$224 \times 224$, which may result in a loss of information for
very large binaries. Conversely, very small binaries present problems too, as upscaling may
introduce interpolation artifacts.
The 17-family evaluation dataset is particularly affected by this, since
the source TIFFs are of varied sizes and must be padded and resized into
the encoder's expected $224\times224$ input before any embedding is
computed.

Our metric learning technique is based on static binary images. It is generally
accepted that static analysis is more efficient than dynamic analysis, but also
more susceptible to elementary obfuscation techniques~\cite{Anusha}.
It is also the case that static analysis is blind to many important features, including
API calls, network traffic, control flow graphs, runtime unpacking, and many more.

As with other learning models, our metric learning model may be susceptible to
adversarial attack. An attacker could, for example, attempt to craft new malware
samples whose images are similar to an existing family in the embedding space,
thereby yielding a misclassification. However, in practice, modifying the binary to
create adversarial images while maintaining the desired functionality
would likely be challenging.

To address some of the limitations above, the following should be considered.
\begin{itemize}
\item Multi-modal fusion --- Combining image embeddings with byte-level entropy features,
opcode sequences, API call patterns, or other relevant features.
\item Dynamic analysis --- Incorporating execution traces or sandbox reports into images.
\item Unpacking detection --- Identify packed samples and applying unpacking before image conversion.
\item Adversarial training --- Training with adversarial examples to improve robustness to adversarial attacks.
\end{itemize}

We note that most of the limitations above---packing/encryption, metamorphic
obfuscation, very small or very large binaries, the static-analysis blind
spot, and adversarial perturbations---are not specific to metric learning;
they affect any image-based malware analysis technique that relies on
byte-level structure. The one limitation that \emph{is} specific to the
metric-learning formulation is the need to maintain and curate a gallery
of labeled examples for the families one wishes to recognize.

Despite the limitations discussed in this section, our experimental results demonstrate
that image-based metric learning has considerable merit at deployment time:
inference is fast (1\,ms per query on GPU), the gallery scales to large
numbers of samples via FAISS, and---unlike a fixed classifier---onboarding
a newly observed family requires only adding its labeled examples to the
gallery rather than retraining.
Metric learning is especially well suited for use
in combination with other feature analysis techniques 
as part of a multi-stage detection strategy.


\section{Conclusion and Future Work}\label{sect:conc}

In this chapter, we presented a comprehensive metric learning framework for malware
classification and evaluated it in three progressively more challenging settings.
In the same-domain setting on MalImg, our triplet-loss ResNet-18 encoder
achieved~94.0\%\ \texttt{retrieval@1}, outperforming both
an SVM baseline~(86\%) and a supervised
ResNet-18 classifier~(91\%). In the standard cross-domain setting, an encoder
trained on MalNet generalized to MalImg with~88.5\%\ \texttt{retrieval@1}, without
any retraining. In the strict zero-shot setting on a held-out~17-family grayscale
dataset, a lightweight CNN with multi-proxy anchor loss achieved~73.1\%\
\texttt{retrieval@1} and~90.5\%\ open-set AUROC---more than~12$\times$ the random
baseline. These results indicate that the learned embeddings capture
visual patterns in malware images that transfer across datasets and, to a lesser
but still useful extent, across entirely unseen families.

Through ablation studies, we validated key design choices, including the
selection of ResNet-18 over ResNet-50 for the same-domain setting, the surprising
finding that a smaller backbone generalizes better in the strict zero-shot setting,
and the choice of embedding dimension, margin, and batch size.
We also acknowledged various limitations---including reliance on byte-level visual
structure (and hence sensitivity to packing and encryption), the use of a
single random seed per experiment (so reported numbers are point estimates
rather than mean and standard deviation over multiple runs),
and the well-known fragility of static analysis to adaptive adversaries---and
suggested ways to mitigate some of these issues.
Our framework includes complete training, evaluation, and deployment tools with
detailed implementation documentation, making it practical for real-world security
applications and educational use. These tools are available from the authors on request.

In future work, we plan to explore some of the issues mentioned at the end of
Section~\ref{sect:lim}. In particular, we plan to consider multi-modal approaches
combining visual embeddings with byte-level entropy features, opcode sequences,
and dynamic analysis traces to improve robustness against obfuscation techniques.
We also plan to repeat all experiments with multiple random seeds to report
confidence intervals. Additionally, we plan to evaluate our metric learning
approach on larger and more diverse malware datasets, including mobile malware
and very recent Windows malware families, to further validate
cross-domain and zero-shot generalization.
We also intend to extend the robustness analysis of Section~\ref{sect:exp} beyond
the random perturbations in Table~\ref{tab:pert} to adversarial perturbations,
that is, perturbations optimized to maximize the retrieval error rate while minimizing
the perturbation magnitude. Of particular interest is generating such perturbations
directly from real malware binaries---grouped by family, including both original and
adversarially modified variants---and converting them to images before applying the
adversarial perturbation, so that the resulting samples remain valid executables.
This adversarial robustness study is a promising direction for a dedicated follow-up paper.

\bibliographystyle{plain}
\bibliography{refs.bib}

\end{document}